\documentclass[aps
,%
prl,%
twocolumn%
]{revtex4}

\usepackage[dvips]{graphicx}
\def\gsim{\buildrel {\textstyle >}\over {_\sim}}
\def\lsim{\buildrel {\textstyle <}\over {_\sim}}

\begin{document}
\title{Numerical study of the ordering of the $\pm J$ {\it XY} spin-glass 
ladder}
\author{Tsukasa Uda$^{1}$, Hajime Yoshino$^{1,2}$ and Hikaru Kawamura$^{1}$}
\affiliation{
$^1$Department of Earth and Space Science, Faculty of Science,
Osaka University, Toyonaka 560-0043,
Japan\\
$^2$Laboratoire de Physique Th{\'e}orique  et Hautes {\'E}nergies, Jussieu, \\
5\`eme {\'e}tage,  Tour 25, 4 Place Jussieu, 75252 Paris Cedex 05, France
\\
}
\date{\today}
\begin{abstract}
The properties of the domain-wall energy 
and of the correlation length are studied 
numerically for the one-dimensional $\pm J$ {\it XY} spin glass  on the
two-leg ladder lattice, focusing on both the spin and the chirality 
degrees of freedom. 
Analytic results obtained by Ney-Nifle {\it et al\/} for the same model
were confirmed for asymptotically large lattices, while the approach to the
asymptotic limit is slow and sometimes even non-monotonic. 
Attention is
called to the occurrence of the $SO(2)$-$Z_2$
decoupling and its masking in spin correlations, the latter reflecting 
the inequality between 
the $SO(2)$ and $Z_2$ exponents. Discussion is given concerning the
behaviors of the higher-dimensional models.
\end{abstract}
\maketitle

\noindent
\S 1  Introduction
\medskip

The domain-wall method, or the stiffness method, is widely used in studying the ordering properties of various spin systems including the spin glass (SG). 
In this method, one computes by some numerical means  the change of the
ground-state energy of finite systems of the linear dimension $L$ under the 
appropriate change of boundary conditions (BCs) imposed on the system
\cite{Banavar,McMillan,BrayMoore1}. This
energy is called a stiffness energy
(or a domain-wall energy), $\Delta E(L)$, which gives a measure of an 
energy scale of low-energy excitations of size $L$.
For large $L$, $\Delta E(L)$ is expected to behave as  a power-law, 
$\Delta E(L)\approx L^\theta$,
$\theta$ being a stiffness exponent.
If $\theta <0$,
the system remains in the disordered state at any nonzero
temperature, whereas if $\theta >0$
the system possesses a finite long-range order 
at low enough temperatures with $T_c>0$.

For complex systems like SGs in which the nature of the ordering is
highly nontrivial,  some fundamental questions still 
remain in this method. First question concerns with the choice of the
set of BCs. In principle, there could be various choices of the set of BCs, 
and the behavior of $\Delta E(L)$ may depend on these choices, 
particularly for small sizes practically accessible in numerical simulations. 
It is not generally clear which set of BCs should be best 
chosen, especially when  
the results depend on the BCs.
Second question concerns with the meaning of the stiffness exponent. 
When the domain-wall energy decreases with $L$, {\it i.e.\/}, the stiffness 
exponent is negative, it is a common practice to relate
the inverse of the stiffness exponent with the correlation-length exponent 
$\nu$ associated with the $T=0$ transition, {\it i.e.\/}, $\nu=1/|\theta |
=1/y$ ($y=|\theta |$)\cite{McMillan,BrayMoore1,Carter}. 
By contrast, 
there were reports which cast doubt on the validity of this  
simple relation \cite{Kawashima}. 
Furthermore, if the model could exhibit more than one 
stiffness exponents depending on the choice of the set of BCs, a question 
immediately arises which stiffness exponent should be chosen to
estimate the correlation-length exponent. 
Since, in the standard 
continuous (second-order) phase transition, 
only one diverging length scale is expected, 
the existence of more than one stiffness exponent poses some problem 
in its interpretation. 

In fact, under certain circumstances, there could be more than one 
distinct diverging length scales, or more than one distinct
correlation-length exponents, at a single continuous transition. 
An example of this may be seen
in {\it chiral transitions\/} possibly realized in certain
frustrated vector spin systems including the {\it XY } SG \cite{KawaAjiro}. 
Frustrated vector spin systems often possesses a chirality degree of
freedom due to the canted spin structure, according as the noncollinear 
(or noncoplanar) structure induced spin frustration is either
right- or lef-handed. Chirality is a pseudo-scalar variable, being 
invariant under global spin rotations [$SO(2)$] but changing sign 
under global spin reflections [$Z_2$]. Recent studies have suggested that
some of such chiral spin systems might 
possibly exhibit a ``spin-chirality decoupling'' phenomenon, where the $Z_2$ 
chirality exhibits an ordering behavior entirely different from the $SO(2)$ 
spin\cite{KawaAjiro}, 
though there still remains some controversy concerning whether such a 
spin-chirality decoupling really 
occurs \cite{KawaAjiro,Kosterlitz,Granato2,LeeYoung}.

In one possible realization of the spin-chirality decoupling, 
the chirality and the spin  exhibit two separate transitions 
at mutually distinct temperatures, whereas, in other possible realization, 
the chirality and the spin  order at the same temperature 
but with the mutually different  spin and chirality 
correlation-length exponents, $\nu_s$ and $\nu_\kappa$. 
Examples of the first class might be the orderings of
the regularly-frustrated two-dimensional (2D) 
{\it XY } model \cite{KawaAjiro} and those of
the three-dimensional (3D) 
$XY$ SG \cite{KawaTane91,Kawa95,KawaLi}. Examples
of the second class the ordering of the 
regularly frustrated  one-dimensional (1D) 
$XY$ model \cite{Horiguchi} and those of the 2D $XY$ SG 
\cite{KawaTane87,KawaTane91}. 
A firmly established example is the case of the regularly-frustrated 1D $XY$ 
model, where it has been  shown  rigorously
that the spin and the chirality order at $T=0$ where 
the chiral correlation length diverges exponentially with $\nu_\kappa =\infty$ 
but the spin correlation length as a power-law with $\nu_s=1$
\cite{Horiguchi}.

More controversial is the nature of the ordering of the {\it XY} SG. 
Some time ago, 
Kawamura and Tanemura made a numerical domain-wall study of 
the {\it XY} SG in 2D and 3D \cite{KawaTane91,Kawa95}. 
These authors introduced various types of BCs to probe the spin and 
the chirality orderings of the model, including the
periodic (P), antiperiodic (AP) and reflecting (R) BCs. In particular, 
the domain-wall energy obtained under
the combination of the P and AP BCs (P/AP), $\Delta E_{{\rm P,AP}}(L)$, 
and the one obtained under the combination of the R and P BCs, 
$\Delta E_c(L)$, apparently yielded mutually different stiffness exponents, 
which were interpreted as associated
with the spin and the chiral correlation-length exponents, respectively.  
These authors observed that, in 2D, both 
$\Delta E_{{\rm P,AP}}(L)$ and $\Delta E_c(L)$ decreased with $L$, 
characterized, respectively, by mutually different stiffness exponents, 
$y_\kappa\simeq 0.5$ and $y_s\simeq 1.0$. This observation was interpreted
as indicating that the chiral correlation length outgrows the spin correlation 
length at the $T=0$ transition of the model, {\it i.e.\/}, $\nu_\kappa  
> \nu_s$ \cite{KawaTane91}. 
These results were corroborated by several
Monte Carlo (MC) simulations on 
the 2D $XY$ SG \cite{KawaTane87,Batrouni,Ray,Bokil,Wengel,Granato1}.
By contrast, on the basis of their domain-wall energy calculation,
Kosterlitz and Akino claimed that the spin and the chiral
correlation-length exponents
were common at the $T=0$ transition \cite{Kosterlitz}.

In 3D, Kawamura and Tanemura observed that $\Delta E_{{\rm P,AP}}(L)$ 
($\Delta E_c(L)$) decreased (increased) with $L$, which
was interpreted as indicating that the chiral-glass transition occurred
at a nonzero temperature, $T_{{\rm CG}}>0$, while the standard SG 
transition occurred only at $T_{{\rm SG}}=0$ \cite{KawaTane91,Kawa95}. 
MC results supporting such a view were also reported \cite{Kawa95,KawaLi}.
Meanwhile, later domain-wall energy calculation by 
Macourt and Grempel suggested that  $\Delta E_{{\rm P,AP}}(L)$ 
might eventually be iterated toward strong coupling for larger $L$ 
and that $T_{{\rm CG}}>T_{{\rm SG}}>0$ \cite{Grempel}. By contrast,
Lee and Young claimed on the basis of their MC simulations
that the spin and the chirality ordered at the same finite
temperature $T_{{\rm CG}}=T_{{\rm SG}}>0$, with a common correlation-length 
exponent \cite{LeeYoung} $\nu_s=\nu_\kappa$. 
Thus, the situation remains quite controversial.

Inspired by the numerical work of Ref.\cite{KawaTane91} on the 
2D and 3D {\it XY} SG, Ney-Nifle, Hilhorst and Moore performed 
an analytic study
of the 1D $XY$ SG ladder with the bond-random $\pm J$ (or binary) 
interaction \cite{Ney}. 
The Villain's action was assumed there. Via the dual transformation,
the model was 
mapped onto the 1D charge Hamiltonian. Since the mapped model was still not 
amenable to the exact treatment, Ney-Nifle {\it et al\/} made further
simplifications, and eventually 
derived several analytic results concerning the domain-wall energies
and correlation lengths. 
They observed that, depending on the type of the applied BCs 
and on whether the total number of frustrated plaquettes is either
even or odd, the domain-wall energy exhibits different behaviors. 
When the sample average is taken over all samples, 
$\Delta E_{{\rm P,AP}}$ is characterized 
by the chiral stiffness exponent $y_\kappa=1.899\cdots $, 
while $\Delta E_c$ 
is characterized by the spin-wave (SW) stiffness exponent $y'_s=1$. 
This is in contrast to the assignment made in Ref.\cite{KawaTane91}
for the 2D and 3D {\it XY} SGs.

If one looks at the spin and chiral correlations of this 1D ladder model, one
sees that the model exhibits the spin-chirality decoupling 
in the sense that there exist two distinct diverging lengths 
at the $T=0$ transition (though Ney-Nifle {\it et al\/} 
apparently stated otherwise),
the one associated with the $Z_2$ chirality and the other associated with 
the $SO(2)$ SW.  The chiral correlation length
$\xi_\kappa$ is characterized by the exponent $\nu_\kappa =1/y_\kappa 
=0.5263\cdots $, 
while  the SW correlation length
$\xi'_s$ is characterized by the SW exponent $\nu'_s=1/y'_s=1$. The full spin 
correlation function is the product of the $Z_2$ part with the 
correlation length $\xi_\kappa \approx T^{-0.526}$ and 
the $SO(2)$ part with the correlation length 
$\xi'_s\approx T^{-1}$. Reflecting the fact that the $Z_2$ chiral
correlation-length exponent happens to be smaller than the $SO(2)$ SW 
correlation-length exponent, 
{\it i.e.\/}, $\nu'_s>\nu_\kappa$,
the full spin correlation function is dominated by the chiral 
exponent $\nu _\kappa$.
 
The analytic result of Ref.\cite{Ney}, though quite plausible, is
not completely rigorous.  
Furthermore, some of the results were obtained for asymptotically
large lattice. In the SG problem, it is sometimes important, and often 
not a trivial matter, 
to elucidate the finite-size effect, {\it e.g.\/},
how large the system must be for exhibiting 
the asymptotic large-lattice behavior.  

Thus, we feel it would be useful to perform  numerical study of 
the $\pm J$ 1D {\it XY} SG ladder in comparison with the analytic work 
of Ref.\cite{Ney}.
In the present paper, we undertake 
such numerical analysis of 
the $\pm J$ $XY$ SG model on two-leg ladder lattices.
The aim of our calculation is threefold.
(i)  We wish to test the validity of the simplifications made in the analytic 
work of Ref.\cite{Ney}.
(ii) We wish to elucidate the nature of the 
finite-size effect in this 1D model. 
(iii) 
We wish to further examine  
the relation between the stiffness exponents and the correlation-length 
exponents in this 1D model.

The following part of the paper is organized as follows.
In \S 2, we introduce the model and  summarize the analytic results of 
Ref.\cite{Ney}. 
In \S 3, we present our 
numerical results of the domain-wall energies. The results are compared with 
those of the analytic work of Ref.\cite{Ney}. Finite-size effects are 
analyzed carefully. In \S 4, we present our 
numerical results of the spin and chiral correlation lengths, 
in comparison with the corresponding 
analytic results of Ref.\cite{Ney}. Relation with
the stiffness exponents and the correlation-length exponents are examined. 
Finally, \S 5 is devoted summary and discussion.

\bigskip
\bigskip\noindent
\S 2. The model and some analytic formula  
\medskip

The model we consider is, firstly, the standard $XY$-SG model  
on the 1D two-leg
ladder lattice with the  binary (or $\pm J$) interaction,
whose Hamiltonian is given by
\begin{equation}
{\cal H}=-\sum_{<ij>} J_{ij}\vec S_i\cdot \vec S_j
        = -\sum_{<ij>} J_{ij} \cos(\theta_i-\theta_j),
\label{eq-cos-hamiltonian}
\end{equation}  
where $\vec S_i=(S_i^x, S_i^y)=(\cos \theta_i, \sin \theta_i)$ ($0\leq \theta_i <2 \pi$) 
is the two-component spin variable at the site
$i$, and the summation is taken over all nearest-neighbor
pairs on the ladder lattice. The site index $i$ may be written as $i=(x,y)$  
with $1\leq x \leq L$ and $1\leq y\leq 2$, where $y=1$ and 2
refer to the first and the second row of the ladder. $J_{ij}$ 
represents the random variable taking either $+1$ or $-1$
with equal probability independently at each bond. The absolute value of the
exchange interaction has been taken to be a unit of energy ($J=1$). 

In the following, we 
impose several types of BCs on the {\it XY}-spin 
variables at the boundary, {\it i.e.\/}, the periodic (P), antiperiodic (AP) 
and reflecting (R) BCs. In these P, AP and R BCs, we impose the relations,
$\vec S_{(L+1,y)}=\vec S_{(1,y)}$,  $\vec S_{(L+1,y)}=-\vec S_{(1,y)}$ and 
$\vec S_{(L+1,y)}=(S_{(1,y)}^x,-S_{(1,y)}^y$),   respectively. 
In the R BC, we reflect the spin at the boundary with respect to the $x$-axis 
in spin space.
  
The local chirality variable at the plaquette $x$, consisting of 
four spins at the sites ($x,1$),($x+1,1$),  ($x+1,2$) and ($x,2$),  
are defined by
\begin{equation}
\kappa_x=\frac{1}{2\sqrt{2}} \sum_{<ij>} 
{\rm sgn}(J_{ij}) \sin (\theta_i-\theta_j),
\end{equation}  
where the summation is taken over four bonds connecting the above four 
sites forming the plaquette. In the ground state of
an isolated frustrated plaquette, it takes a value either $+1$ or $-1$, while in the ground state of an isolated unfrustrated
plaquette, it takes a value equal to zero. Thus, the 
states with $\kappa =\pm 1$
represent the two chiral states, with right-handed and left-handed spin
circulation around the plaquette. 

Other model we consider is the effective charge Hamiltonian on the dual
lattice. The simplest version is the so-called Villain's Hamiltonian, which contains only the 2-body charge interaction. Although it has commonly been believed that the Villain's Hamiltonian 
becomes equivalent to the cosine Hamiltonian in the low-temperature 
limit $T\rightarrow 0$ \cite{Ney}, we have found that this is actually not the case: The original {\it XY\/} Hamiltonian mapped to the charge representation contains the higher-body interactions, in addition to the 2-body interaction, even in the $T\rightarrow 0$  limit. Thus,
in the present work, we also consider these higher-body correction terms 
to the standard Villain's 2-body approximation \cite{unpublished}. The explicit forms are given
in the appendix.

In the cases of the P and AP BCs, the two-body charge Hamiltonian, or the Villain's Hamiltonian, takes the form,
\begin{equation}
{\cal H}_{{\rm P}}= \sum_{i,j}U_{ij} m_im_j 
         + \frac{\pi^2}{L} (2n- \sum_i m_i - {\cal P})^2,
\end{equation} 
\begin{equation}
{\cal H}_{{\rm AP}}= \sum_{i,j}U_{ij} m_im_j 
         + \frac{\pi^2}{L} (2n- \sum_i m_i - {\cal P} + 1)^2,
\end{equation} 
respectively,
where the charge variable $m_i$, sitting at  
the plaquette $i$, takes integer values $0,\pm1,\pm2 ...$ on
unfrustrated plaquettes, and half-integer values $\pm \frac{1}{2}, \pm \frac{3}{2} ...$ on frustrated plaquettes.
The variable $n$ takes integer values 
$0,\pm1,\pm2 ...$, while ${\cal P}$ is the ``parity'' variable 
being equal to zero 
or unity depending on whether the total number of antiferromagnetic
bonds on the first row ($y=1$) of the ladder is either even or odd. 
The interaction between the charge variables
$U_{ij}$ located at the plaquettes $i$ and $j$ is defined by
\begin{equation}
U_{ij}=\frac{\pi^2}{L}\sum _k \frac{e^{ik(i-j)}}{2-\cos k}.
\label{eq-uij}
\end{equation} 
where the summation over the 
wavevector $k$ is taken over $k=0, \pm \frac{2\pi}{L},\pm \frac{4\pi}{L}, \cdots$.
In the $L\rightarrow \infty$ limit, $U_{ij}$ reduces to
\begin{equation}
U_{ij}=\frac{\pi^2}{\sqrt{3}}(2-\sqrt{3})^{|i-j|}, 
\label{eq-u}
\end{equation}
which decays exponentially with distance $|i-j|$.
The first term of Eq.(2) and (4) represents the charge-charge interaction, 
while the second term of Eq.(3) and (4) represents the SW term, which is related
to the charge part via the total charge $\sum _im_i$.

In the case of the R BC, by contrast,
the corresponding two-body charge Hamiltonian is given by
\begin{equation}
{\cal H}_{\rm R} = \sum_{ij} U_{ij} m_im_j,
\end{equation} 
where $U_{ij}$ is still given by Eq.(5), but the summation over the 
wavevector $k$ is now taken over $k=\pm \frac{\pi}{L},\ 
\pm \frac{3\pi}{L}, \cdots$ which yields
$U_{i+L,j}=-U_{i,j}$. Note that there is no 
second term (SW term) in ${\cal H}_{\rm R}$.
 
To proceed further, Ney-Nifle {\it et al\/} made the following two 
assumptions \cite{Ney}. First, the charge variable $m_i$ is restricted to
$\pm \frac{1}{2}$ on  frustrated plaquettes and $0$ on unfrustrated 
plaquettes. If one labels the frustrated plaquettes as $I=1,2,\ldots N_{\rm fr}$
, where $N_{\rm fr}$ is the total number of frustrated plaquettes,
the chiral part of the Hamiltonian 
reduces to the 1D Ising Hamiltonian with $N_{\rm fr}$
Ising variables $\sigma_{I}=\pm 1$.
Second, Ney-Nifle made a further simplification that the charge-charge
interaction, which originally work between arbitrary pairs of 
frustrated plaquettes, is restricted only to the nearest-neighbor pairs
of frustrated plaquettes.
After these two simplifications,
the model reduces to the 1D Ising chain with the random 
antiferromagnetic nearest-neighbor 
interaction,
\begin{equation}
{\cal H_{{\rm Ising}}}=\sum _{I=1}^{N_{\rm fr}} V_{I} \sigma_I \sigma_{I+1}. 
\end{equation}
The random nearest-neighbor interaction $V_{I} >0$ obeys the distribution
given by
\begin{equation}
P(V)=cV^{-1+\nu_\kappa},\ \ \ \nu_\kappa=0.5263\cdots ,
\label{eq-pv}
\end{equation}
for smaller $V$, 
where $c$ is a normalization constant. The latter simply follows from
Eq.(\ref{eq-u}) and the fact that the probability
to have a sequence of $l$ successive unfrustrated plaquettes
is given by  $1/2^{l}$. Thus, it should be remarked that
the distribution of the effective interactions $P(V)$ is not
a smooth function but a collection of delta functions 
so that Eq.(\ref{eq-pv}) must be taken with care.

These simplifications enabled Ney-Nifle {\it et al\/} to specify
the explicit charge (chirality) pattern and the existence/non-existence of the 
SW excitation in the
ground state of large enough lattices under the given BC, leading to
various predictions on the domain-wall energies and the correlation lengths. 

\bigskip
\bigskip
\noindent
\S 3. Numerical results on the domain-wall energies
\medskip

In this section, we numerically calculate the following two types of
domain-wall energies 
for the 1D {\it XY} SG ladder, {\it i.e.\/}, 
(i) the root-mean square of the energy 
difference between under the P and AP BCs, $\Delta E_{{\rm P,AP}}$, 
and (ii) the absolute value of the energy difference 
between under the R and min(P,AP) BCs, $\Delta E_{{\rm c}}$, where
min(P,AP) refers to either P or AP BC which has the lower energy than 
the other.    
The following three levels of numerical calculations are made.

\noindent
{\it Method (A)}: The first method is the direct numerical estimate of 
the ground-state energy of the cosine Hamiltonian of finite $L$ 
under the given BC.
In calculating the ground-state energy, we employ the spin-quench
algorithm without any further approximation, {\it i.e.\/}, by starting from the randomly generated spin initial
conditions, we quench the system to reach one of the local energy minima. 
These quench procedures are repeated
many times, typically 5,000 times, 
until one is sure that the true ground state has 
been reached. If one goes to
larger $L$, the number of local minima increases rapidly
which makes the search of the true ground state
increasing difficult. This difficulty limits the tractable maximum lattice 
size to $L\leq 35$. The sample average is taken over 10,000 ($L=35$) -
200,000 ($L\leq 15$) independent bond realizations.
 
\noindent
{\it Methods (B1)-(B3)}: 
In this group of methods, we estimate the ground-state energy of the 
effective charge Hamiltonian of finite $L$ under the
given BC within the first approximation of Ref.\cite{Ney}. Namely,
we use the Ising Hamiltonian but with the distant-neighbor interaction
in estimating the domain-wall energy. 
In identifying the chirality pattern of the ground state (or the
candidate of the ground state), we have made certain plausible
assumptions, the detail of which will be given below for each different
case: It is similar to the  procedure employed in
Ref.\cite{Ney}, but it is systematically improved in the estimation of the
domain wall energies by taking into account the distant-neighbour two-body
interactions and also the higher-body intereactions.
By the methods (B1)-(B3), we examined lattices considerably larger than the case (A), up to $L=960$.
The sample average is taken over 100,000 independent bond realizations.

\noindent
{\it Method (C)}: In this 
third method, we estimate the ground-state energy of the 
Villain's Hamiltonian of finite $L$ under the
given BC  by assuming both the first and the second 
approximations of Ref.\cite{Ney}. Namely,
we use the Ising Hamiltonian with the nearest-neighbor interaction
in estimating the domain-wall energy.  
In this case, we can deal with lattices still larger than the cases (B1)-(B3), and 
the results for asymptotically large $L$ should reduce to
the analytic results of Ref.\cite{Ney}.
The sample average is taken over 100,000 independent bond realizations.

We compare the results of these five levels of calculations 
(A), (B1)-(B3) and (C)  
to examine the validity of
the approximations made in Ref.\cite{Ney}, and to elucidate the nature of
the finite-size effects in this model.

In the original 
cosine model (1), the local chirality at each plaquette is given by 
Eq.(2). In the ground-state, 
the {\it local\/} chirality distribution is expected to be peaked 
around $\kappa=\pm 1$ for frustrated plaquettes,
and around  $\kappa=0$ for unfrustrated plaquettes. 
We have checked  numerically that
this is indeed the case. As shown in Fig.1, the local chirality on 
frustrated plaquettes takes the values around
$\kappa \simeq \pm 1$ with equal probability, while that 
on unfrustrated plaquettes takes the values around
$\kappa = 0$. Such a distribution of the local chirality enables us to  
label the chirality pattern uniquely only 
by the combination of $+$ and $-$ on frustrated plaquettes. 
\begin{figure}[ht]
\begin{center}
\includegraphics[scale=0.7]{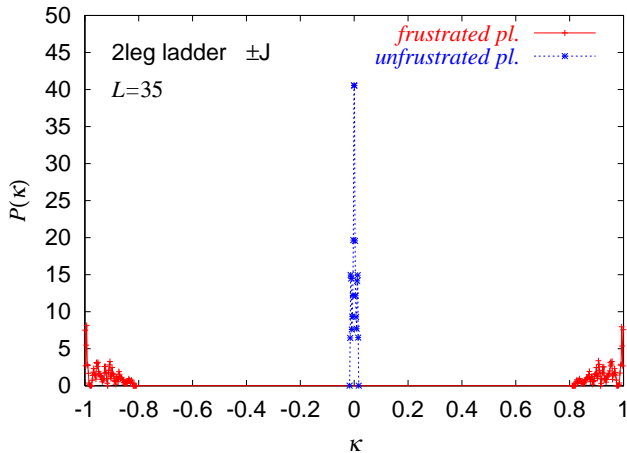}
\end{center}
\caption{
The distribution function of the local chirality $\kappa$ at each plaquette
in the ground state of the $\pm J$ $XY$ ladder. The lattice size is 
size $L=35$. The data for frustrated and 
unfrustrated plaquettes are given in blue (at the center $\kappa=0$)
and in red (near the edges $\kappa=\pm 1$), respectively. 
}
\end{figure}

In the present 1D model,  as was first 
suggested by Ney-Nifle {\it et al\/} \cite{Ney},
the behavior of the domain-wall energy largely depends on whether the 
total number of frustrated plaquettes $N_{\rm fr}$ is either even or odd. 
We call these samples ``even'' and ``odd'' samples, respectively.
In what follows, we show the results for the 
even and odd cases separately.

\bigskip
\bigskip
\noindent
\S 3.1  Even samples
 
\bigskip
\noindent
3.1.1 The domain-wall energy: $\Delta E_{{\rm P,AP}}$

For even samples, Ref.\cite{Ney} predicted that the domain-wall energy 
$\Delta E_{{\rm P,AP}}$
was dominated for large enough lattices by the contribution of 
a pair of chiral domain-wall excitations 
not accompanying 
the SW excitation. The chiral domain-wall may be defined here as the place 
where the ``chiral overlap'' between the two chirality configurations 1 and 2
under the two BCs, $O_i=m_i(1)m_i(2)$, changes the sign. 
According to Ref.\cite{Ney}, 
under the min(P,AP) BC, the sign of the chirality pattern alternates on 
frustrated plaquettes without misfit, while under the max(P,AP) BC,
a pair of misfits is introduced into the alternating chirality pattern, but
not accompanying the SW excitation:  A pair 
of chirality misfits is introduced into the sample 
in such a way that one is at the weakest connection, {\it i.e.\/},
the place where the neighboring frustrated plaquettes are most far apart in 
distance, and the other is at the next-weakest connection satisfying the
condition $\sum_i m_i=\pm 1$. The latter condition is required to suppress
the SW term in eqs.(3) or (4). 
The other possible candidate of the ground state 
under the max(P,AP) BC might be the
$\pi$-SW state with a nonzero second term but without any misfit in
the alternating chirality pattern.
However, if the assumptions made in Ref.\cite{Ney} are to be justified, 
the chiral domain-wall state always has the lower energy than the SW state, 
at least in sufficiently large lattices. 

In our methods (B1)-(B3) above, we search for the positions of a pair of chiral
domain walls in the ground state under the max(P,AP) BC according to the
following procedure. First, to specify the position of one of the two
chiral domain walls, we apply  the R BC to the same sample. The
application of the R BC is expected to yield a single chiral domain wall
in the sample: See below. By calculating and comparing the energies
corresponding to all possible $N_{\rm fr}$ positions of a chiral domain
wall, we determine the position of the chiral domain wall in the ground
state under the R BC.  This position is assumed to be common with the 
position of one of the two chiral domain walls introduced under the
max(P,AP) BC.  We then determine the position of the second chiral
domain wall in the ground state under the max(P,AP) BC by calculating
and comparing the energies corresponding to all possible 
positions of the second chiral domain wall under the constraint that
there are odd number of frustrated plaquettes
between the two chiral domain walls. In calculating the energy, we 
employ systematically imporved methods (B1)(B2) and (B3) unlike the case of our method (C)
which takes into account the nearest-neighbour intereaction only.
In the method (B1), we use the two-body approximation but sum over all
distant-neighbor interactions. In the methods (B2) and (B3), we take into
account the higher-body correction terms up to the 4-body and 6-body
interactions, respectively. The explicit forms of the higher-body terms
are given in the appendix. 

In our direct method (the method (A) above) 
for finite $L\leq 35$ samples, we have observed that, under the min(P,AP) BC, 
the  rule of the ground-state configuration of \cite{Ney} is always satisfied, while,
under the max(P,AP) BC, 
some samples obey the rule of Ref \cite{Ney}, but some other samples do not.
In the latter class of samples, the ground state under 
the max(P,AP) BC
turns out to be the $\pi$-SW state rather than the chiral
domain-wall state, {\it i.e.\/},
the chirality pattern completely alternate without misfit while the SW
of a turn angle $\pi$ ($\pi$-SW) is generated between under the P and AP BCs. 
An example of such a $\pi$-SW sample is
shown in Fig.2, where the spin configurations under the P and AP BCs
are shown in the upper panel of Fig.2, while the relative
deviation angle between the spin directions under 
the P and AP BCs is illustrated 
in the lower panel by arrows. The appearance of the $\pi$-SW is
clearly visible here. Hence, at least in a subset of samples of finite 
$L\leq 35$, 
the rule of Ref.\cite{Ney} is violated.
\begin{figure}[ht]
\begin{center}
\includegraphics[scale=0.7]{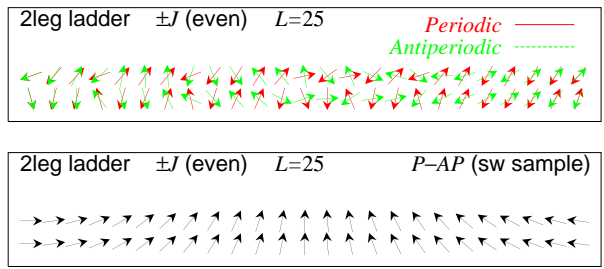}
\end{center}
\caption{
A typical example of the ground-state spin configuration of even samples
under the periodic
and antiperiodic boundary conditions, between which a spin-wave of a turn 
angle $\pi$ is generated. The lattice size is $L=25$.
In the lower panel, the relative
deviation angle between the spin directions under 
the periodic and antiperiodic boundary conditions is illustrated by arrows.
}
\end{figure}

The domain-wall energy $\Delta E_{{\rm P,AP}}$ calculated in our direct 
method (A) is shown in Fig.3(a) on a log-log plot.
The slope is estimated to be about 1.39  in the range $L\leq 35$, 
which is considerably smaller than the 
predicted value of Ref.\cite{Ney}, $1.899\cdots $,
presumably due to the existence of the 
SW samples characterized by the SW stiffness exponent $y'_s=1$. 

\begin{figure}[ht]
\begin{center}
\includegraphics[scale=0.5]{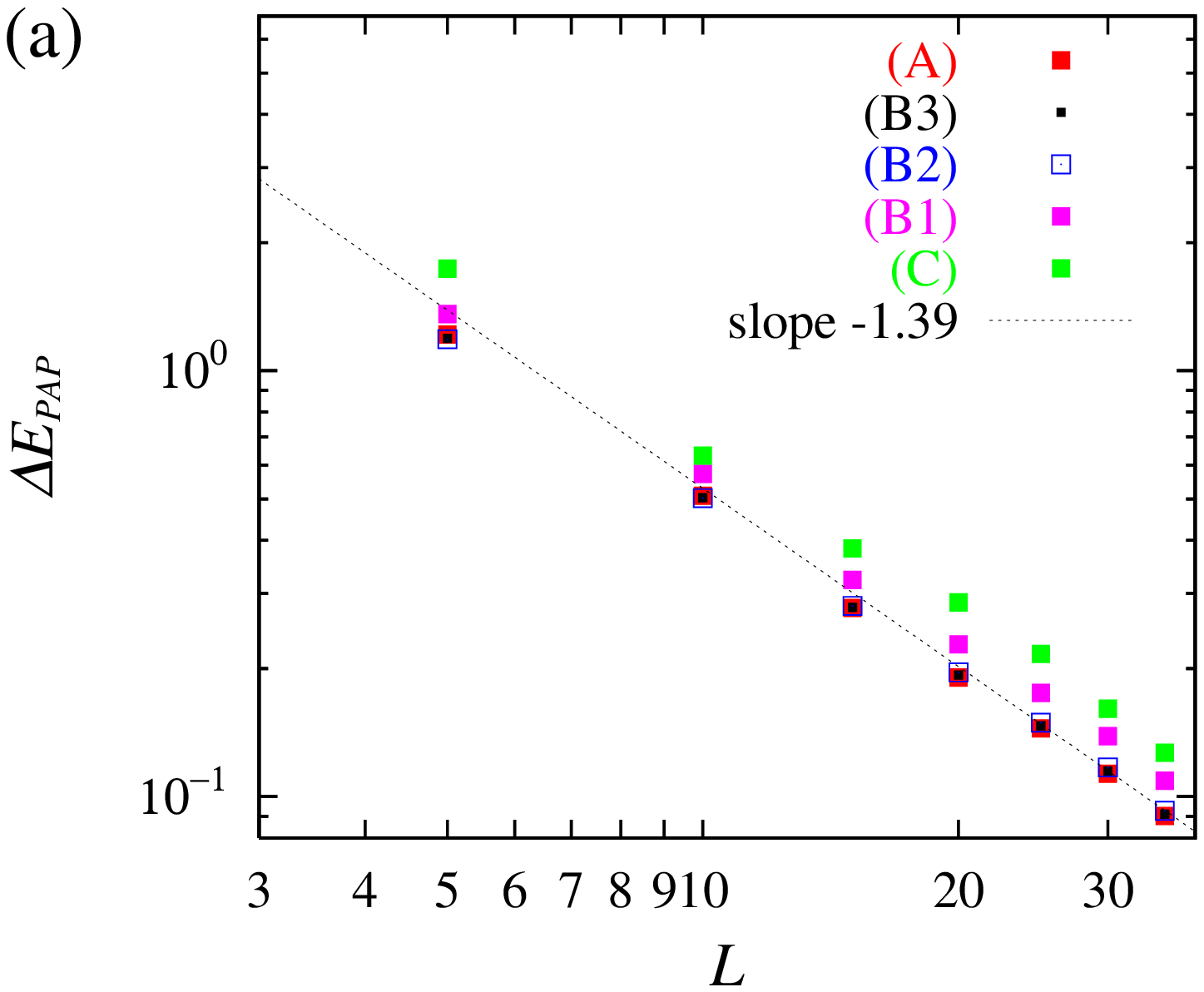}
\includegraphics[scale=0.5]{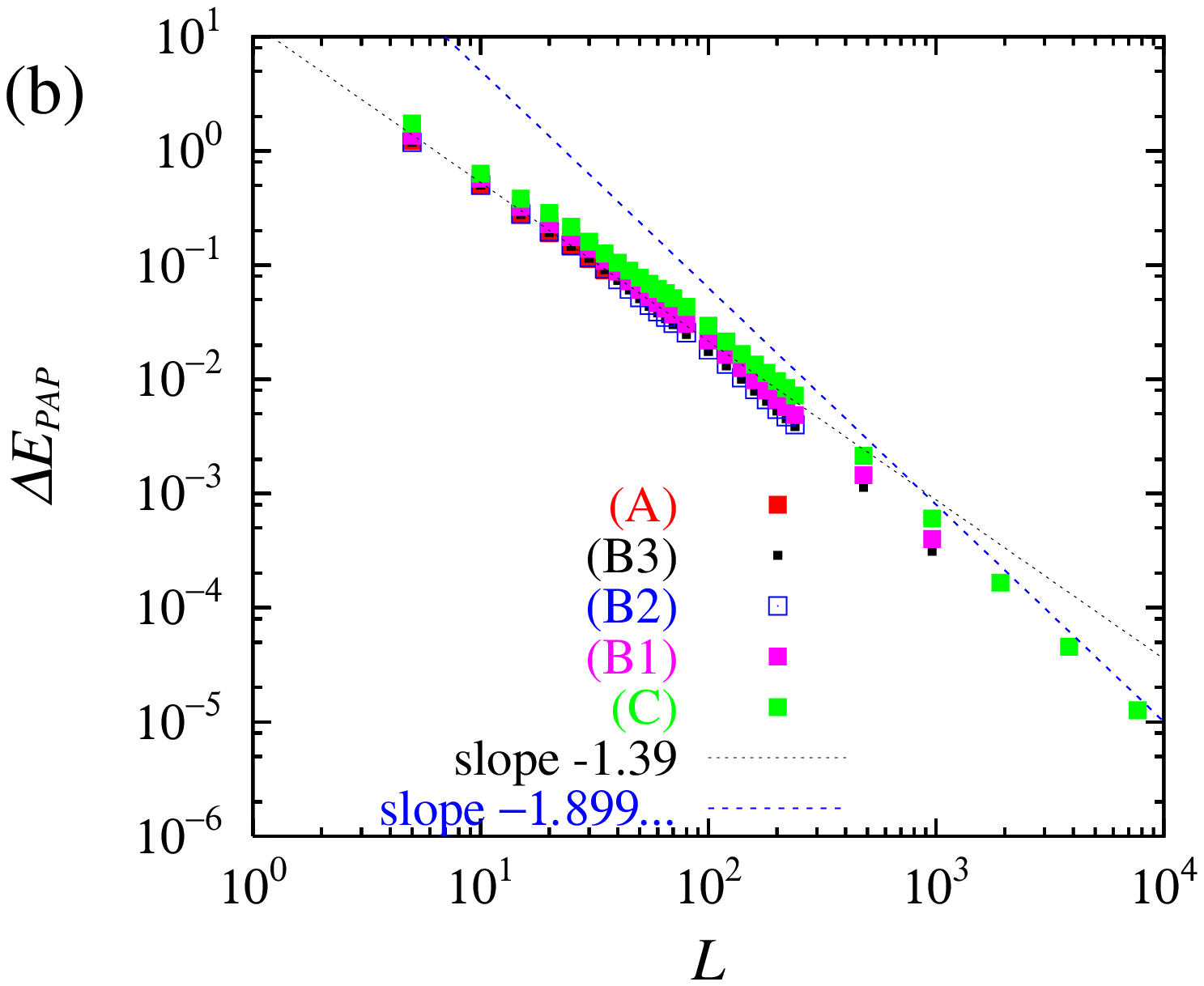}
\end{center}
\caption{
The domain-wall energy $\Delta E_{{\rm P,AP}}$ of even samples,
calculated by the three
methods (A), (B1)-(B3) and (C) mentioned in the text, are
plotted versus $L$ on a log-log plot. 
The dotted line is the  power law $L^{-1.899...}$ predicted in
 Ref. \cite{Ney}.
}
\end{figure}

Then, the next question is how the rate of the $\pi$-SW samples, which breaks 
the rule of Ref.\cite{Ney}, varies with increasing $L$.  
This rate of the SW samples $r_{\rm SW}$
calculated in the direct method (A) is shown in Fig.4(a)
in the range $L\leq 35$. As can be seen from figure, while the rate
decreases with increasing $L$ 
for smaller lattices of $L\lsim 20$, it tends to {\it increase again\/}
for larger lattices up to $L=35$. 
Unfortunately, the direct calculation is limited to $L=35$.

\begin{figure}[ht]
\begin{center}
\includegraphics[scale=0.5]{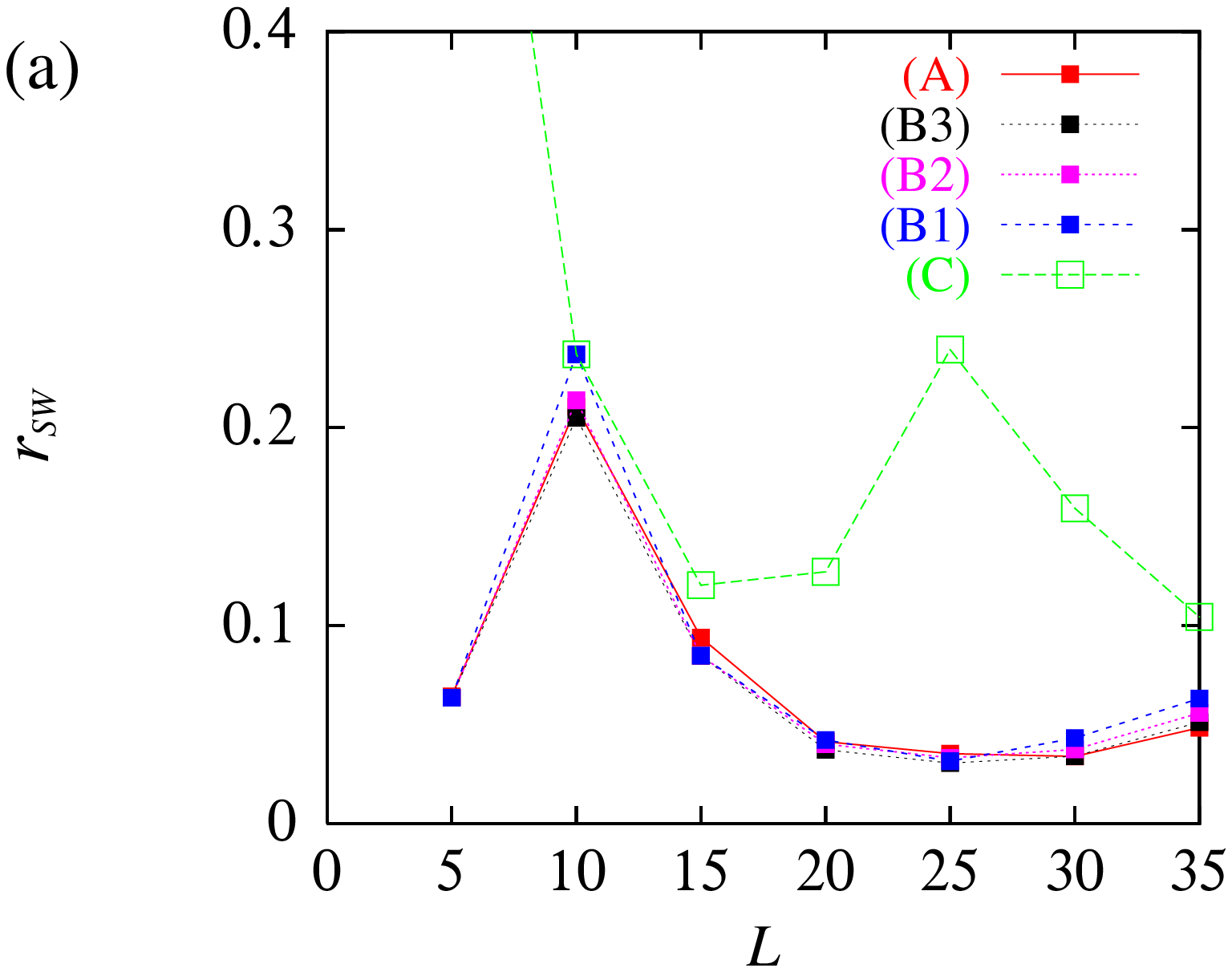}
\includegraphics[scale=0.5]{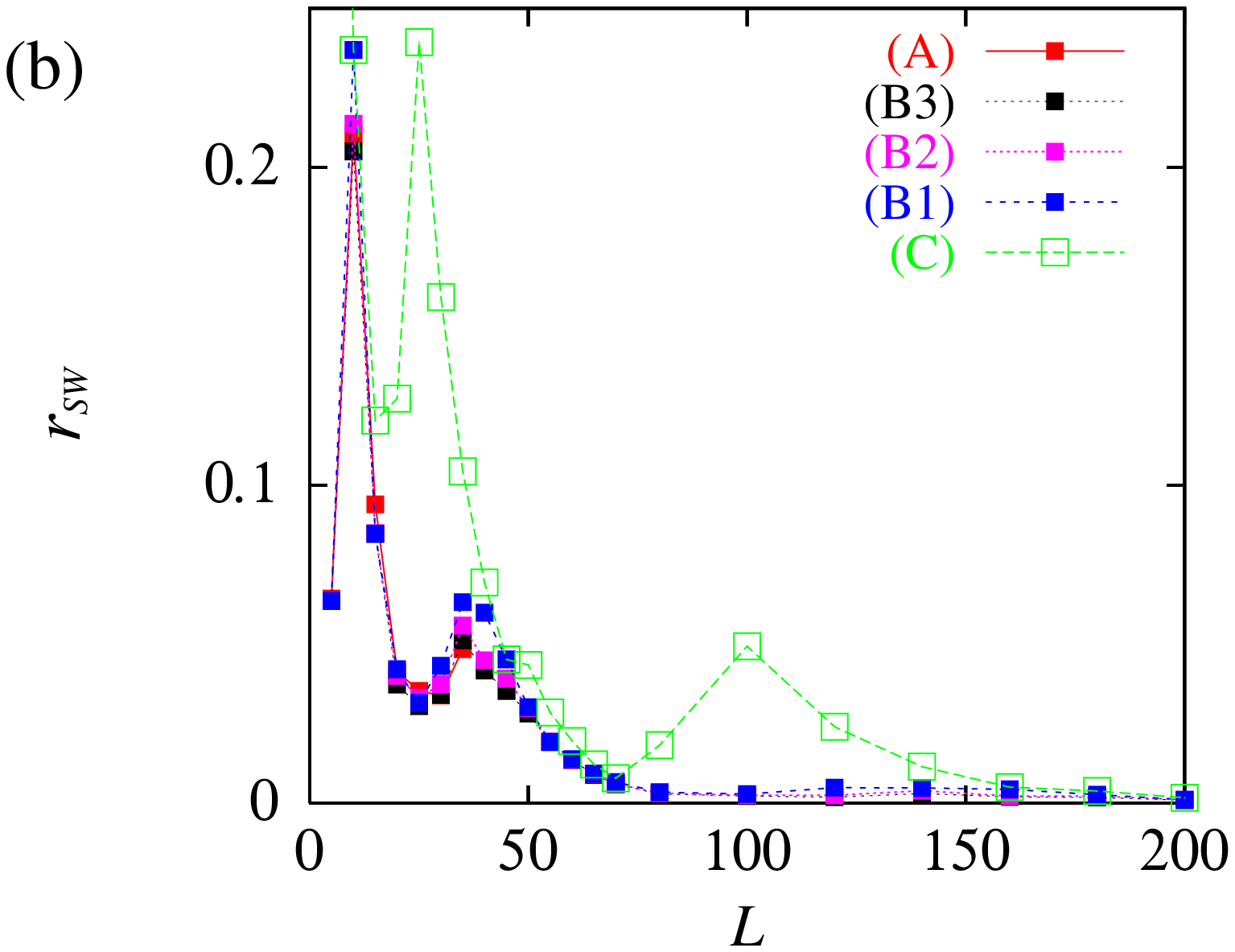}
\end{center}
\caption{The $L$-dependence of the rate of the spin-wave samples 
within the even samples under the periodic
boundary condition, calculated by the 
three methods (A), (B1)-(B3) and (C) mentioned in the text. 
}
\end{figure}

The rate of the SW samples $r_{\rm SW}$ is also estimated
by using the Ising approximation (the method (B1)-(B3)) and the nearest-neighbor
approximation (the method (C)), and the results 
are also shown in Fig.4(a) in the same range of 
$L\leq 35$. 
One sees from the figure that
the rate calculated by both approximations exhibits the non-monotonic behavior qualitatively 
similar to the one observed by the direct method (A). In particular,
the Ising approximation (B1)-(B3) gives the results in quantitative agreement 
with those of the the direct method (A). The quantitative agreement
becomes systematically better by including the distant-neighbour two-body
interactions (B1), 4-body interactions (B2) and 6-body interactions (B3).

By contrast, the nearest-neighbor approximation 
(C) yields the results which considerably deviates 
from the results of the direct method quantitatively, although
some qualitative features are still captured. 

In order to investigate the behavior of larger lattices $L\geq 35$, we have to 
rely on the approximate methods (B1)-(B3) and (C). 
The domain-wall energy $\Delta E_{{\rm P,AP}}$ and the rate of the SW sample
$r_{\rm SW}$ calculated for larger lattices $L\geq 35$ 
by the methods (B1)-(B3) and (C) are shown in Figs.3(b) and 4(b), respectively. 
In both methods, the ground state under the given BC
is searched for between the chiral domain-wall state and the $\pi$-SW state by comparing the energies of 
these two states.
As can be seen from Fig.4(b), the rate of the SW sample $r_{\rm SW}$, once increased 
with $L$ at $L\leq 35$, decreases 
with further increasing $L$ and tends to zero in the $L\rightarrow \infty$ limit 
but with peculiar oscillations. Likewise, as can be seen
from Fig.3(b), $\Delta E_{{\rm P,AP}}$ yields a slope close to 
$1.9$ for large enough $L$, consistent with the value
of Ref.\cite{Ney}. Thus, the asymptotic large-$L$ behavior seems 
consistent with Ref.\cite{Ney}, while the approach to the large-$L$ 
asymptote is rather slow,
realized only for lattices with $L\gsim 40$.

Of course, the methods (B1)-(B3) and (C) assume properties of the
ground-state configurations are not completely rigorous.
However, the fact that
the non-trivial (non-monotonic) small-$L$ behavior revealed by the direct 
method (A) is also reproduced by these
approximate methods gives some credence to the reliability of the 
approximate methods
even for larger $L$ where the direct method is not available. 

In Fig.5, we show the distribution function of the domain-wall energy 
$\Delta E_{{\rm P,AP}}$ 
for the sizes of $L=35$, Fig. 5 (a), and $L=240-960$, Fig. 5 (b).
As can be seen from  Fig. 5(a), 
the SW samples contributes the component near the edge of the
distribution whose weight decreases with increasing $L$. Interestingly,
the distribution is not smooth at all. 
As $L$ increases, more bands of spikes appear closer to the center
$\Delta E_{\rm P,AP}=0$ and the amplitude of the spikes at smaller 
$|\Delta E_{\rm P,AP}|$ becomes larger. Presumably the bands of the spikes 
reflect nearly discrete spectrum of the distribution of
effective interactions between chiralities, which is not explicit in Eq. (\ref{eq-pv}).

\begin{figure}[ht]
\begin{center}
\includegraphics[scale=0.5]{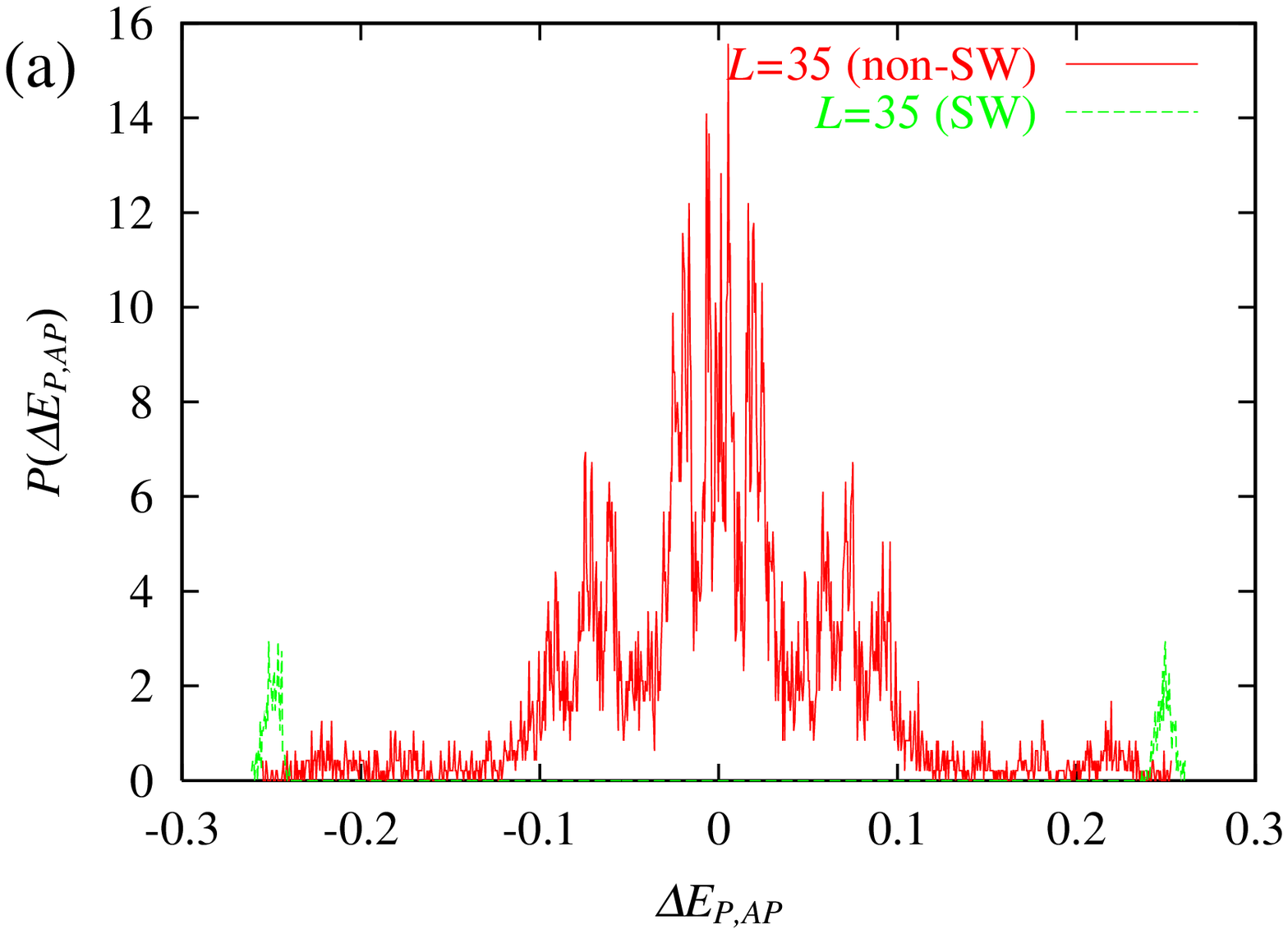}
\includegraphics[scale=0.5]{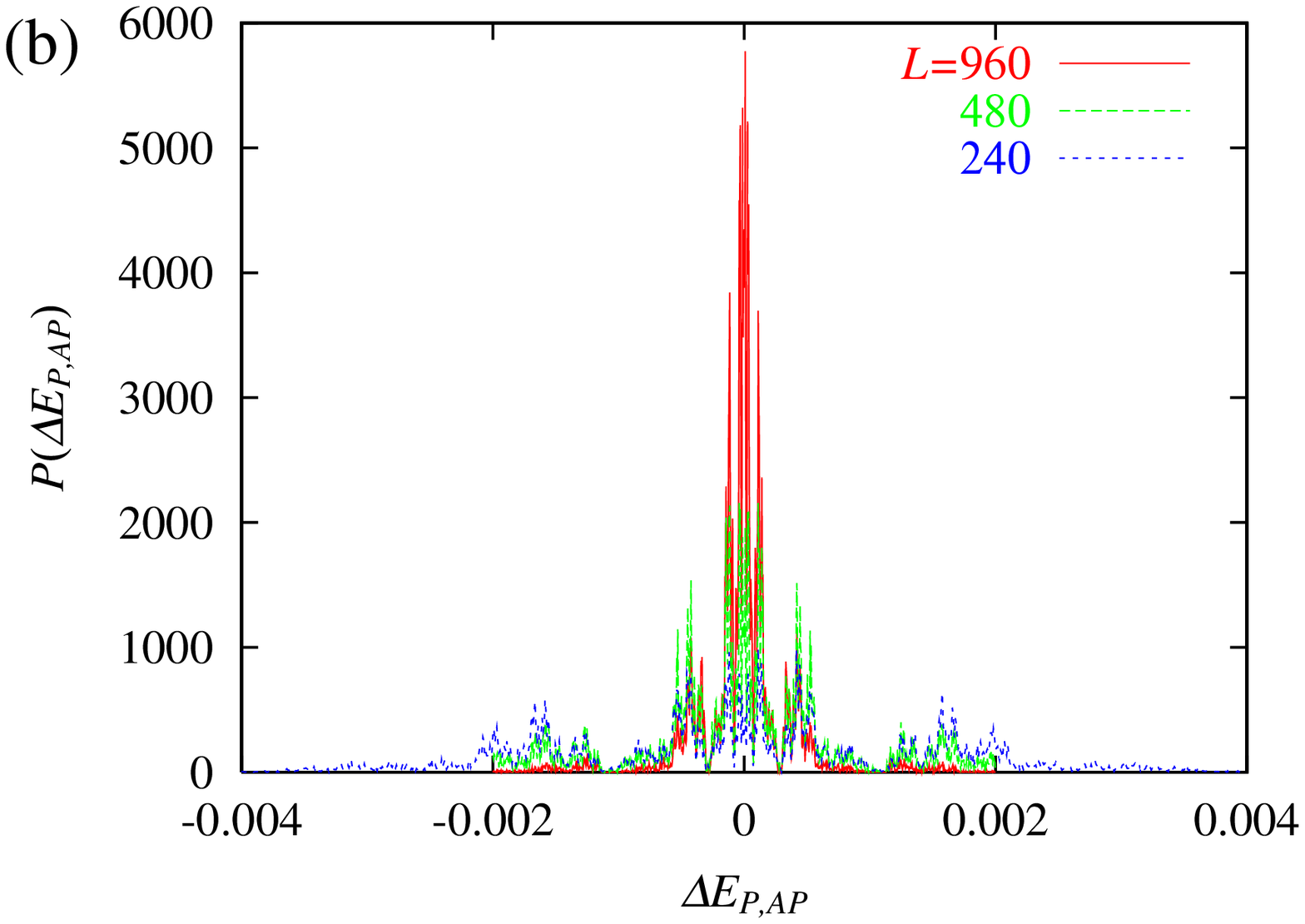}
\end{center}
\caption{
The distribution function of the domain-wall energy $\Delta E_{{\rm P,AP}}$ for
even samples with $L=35$ obtained by the method (A) (a), and $L=240,480,960$
 obtained by the method (B1) (b).
}
\end{figure}

\bigskip
\noindent
3.1.2 The domain-wall energy: $\Delta E_c$

Now, we turn to the second type of domain-wall energy,  
$\Delta E_c$, the absolute value
of the ground-state energy difference
between under the R and min(P,AP) BCs. 
Under the R BC, the sign of the chirality is reversed at the boundary.
Therefore, if
the sign change in the chiral-overlap 
$O_i=m_i({\rm R})m_i({\rm min(P,AP)})$ occurs at the boundary, it actually 
means that there is no
chiral domain-wall at the boundary.
With this understanding, 
under R/min(P,AP) of even samples, 
a single chiral domain-wall should be introduced
into the sample, not accompanying the SW excitation\cite{Ney}. 
Hence,  $\Delta E_c$ should be characterized by the chiral stiffness exponent 
$y_\kappa =1.899\cdots $. 
Indeed, our direct method (A) has fully confirmed this expectation. 

\begin{figure}[ht]
\begin{center}
\includegraphics[scale=0.5]{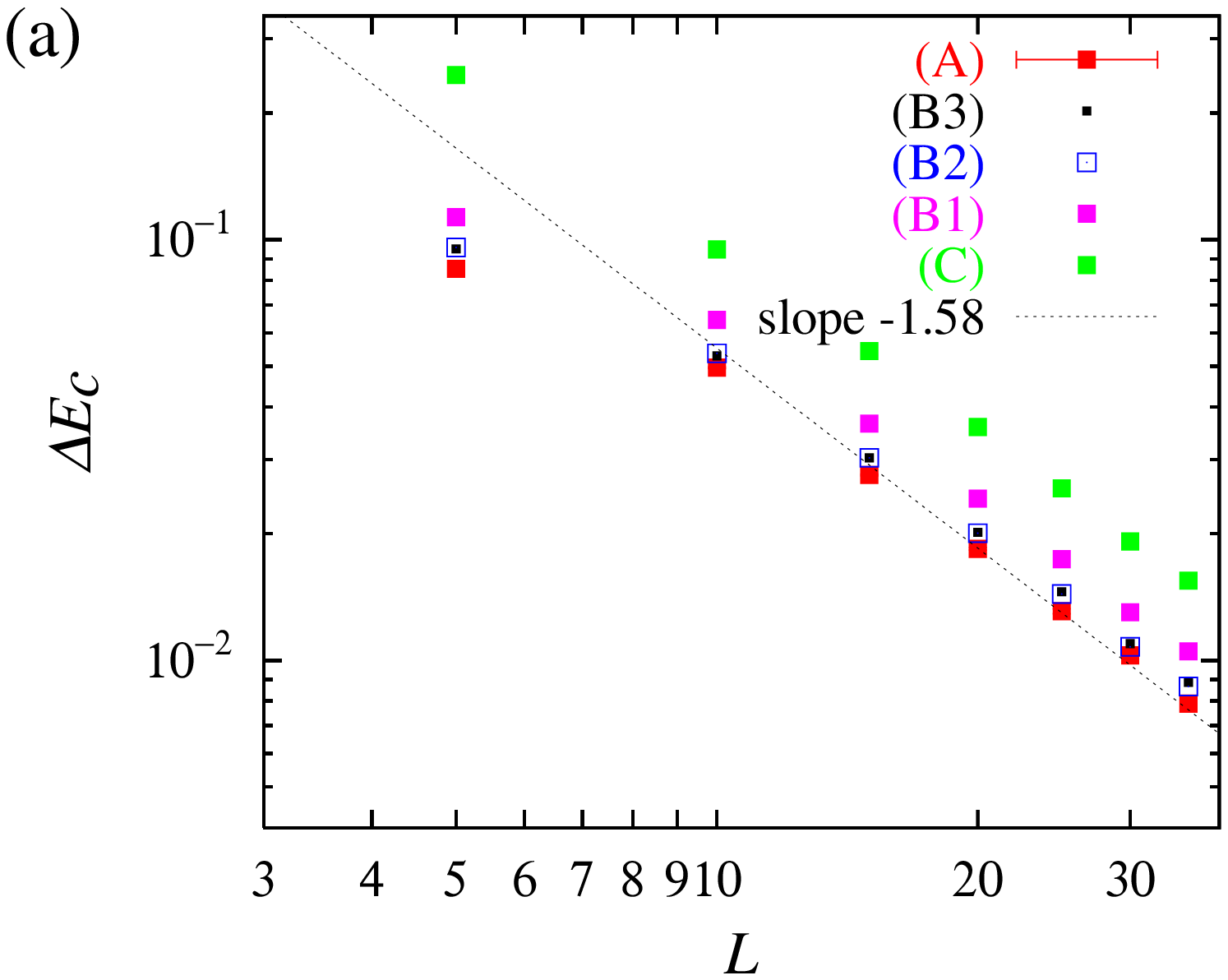}
\includegraphics[scale=0.5]{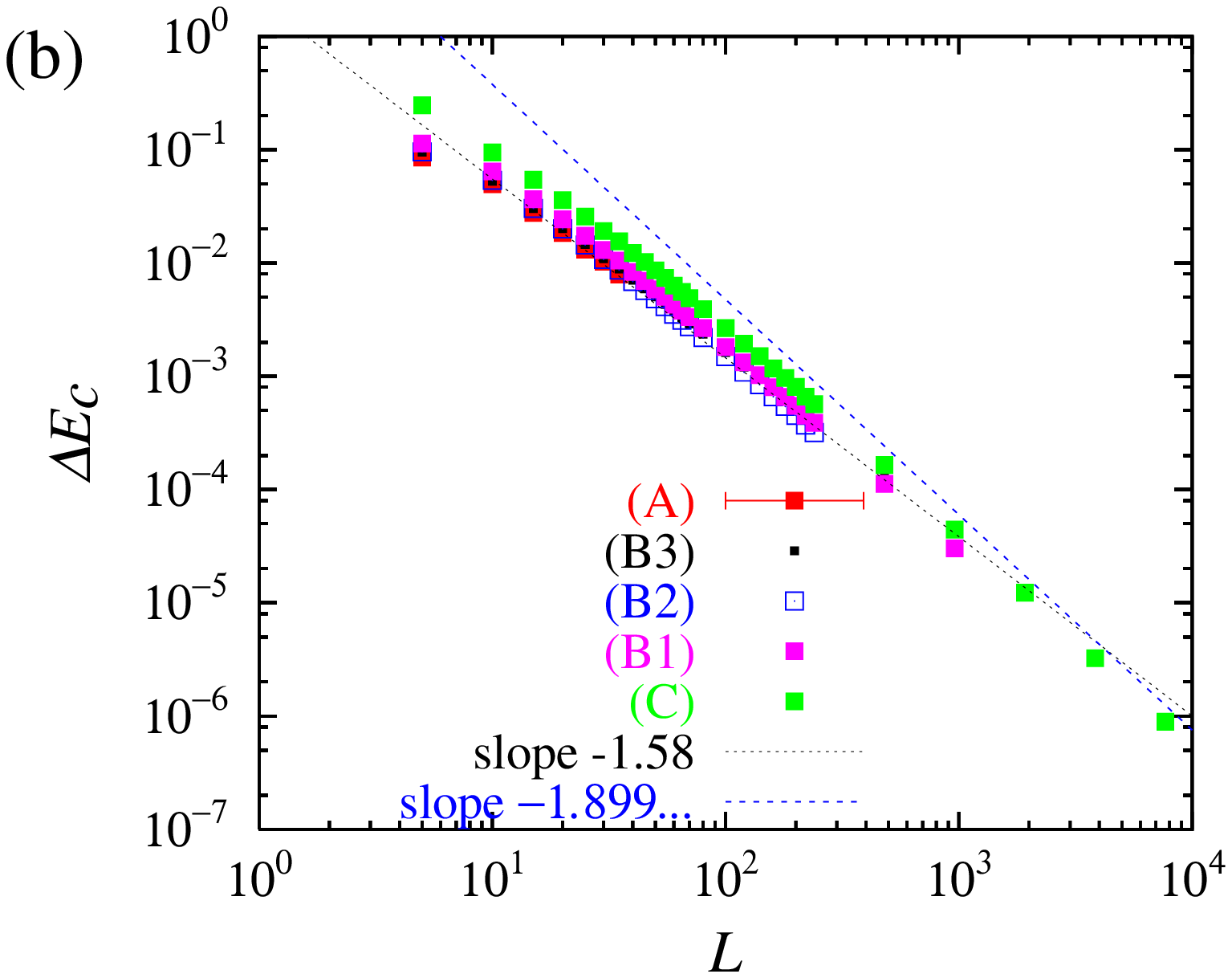}
\end{center}
\caption{
The domain-wall energy $\Delta E_c$ of even samples,
calculated by the three
methods (A), (B1)-(B3) and (C) mentioned in the text, are
plotted versus $L$ on a log-log plot. 
The dotted line is the  power law $L^{-1.899...}$ predicted in
 Ref. \cite{Ney}.
}
\end{figure}

In Fig.6(a), we show the size dependence of the domain-wall energy, 
$\Delta E_c$, estimated by the
three methods (A), (B1)-(B3) and (C). 
As mentioned in \S 3.1.1, in the method (B1)-(B3), the position of a chiral
domain wall is determined by calculating and comparing the energies
corresponding to all possible $N_{\rm fr}$ positions of a chiral domain
wall. In the size range $L\leq 35$ where the direct 
calculation is available,
the data  yield a slope about 1.58, 
which is considerably smaller than the expected 
value $y_\kappa =1.899\cdots $.
However, the data for larger lattices obtained by the approximate methods (B1)-(B3) 
and (C) yield
an asymptotic slope consistent with the expected value $y_\kappa 
=1.899\cdots $. 
Again, the approach to the asymptotic behavior turns out to be
rather slow.

\bigskip
\bigskip
\noindent
3.2  Odd samples

In this subsection, we deal with the other subset of samples where 
the total number of frustrated plaquettes is odd.

\bigskip
\noindent
3.2.1 The domain-wall energy: $\Delta E_{{\rm P,AP}}$

For odd samples, Ref.\cite{Ney} shows that 
the P and AP BCs always yield exactly the same
ground-state energy, {\it i.e.\/}, $\Delta E_{{\rm P,AP}}=0$. 
In the ground-state chirality pattern, 
a single misfit is introduced into the alternating chirality pattern,
but always in the same position between under the  P and
AP BCs so that there is no chiral domain-wall between 
under the P and AP
BCs. Instead, the $\pi$-SW is generated between them. Indeed,
we have confirmed this expectation by the direct method (A). 

\bigskip
\noindent
3.2.2 The domain-wall energy: $\Delta E_c$

In the case of R/min(P,AP) of odd samples, a single chiral 
domain-wall is expected to be introduced with 
accompanying the $\pi/2$-SW\cite{Ney}.  This has been confirmed by our 
direct method (A). A typical
example of the spin configurations under the R and  
P BCs is shown in Fig.7. The existence of the
$\pi/2-$ SW is clearly visible. 

\begin{figure}[ht]
\begin{center}
\includegraphics[scale=0.7]{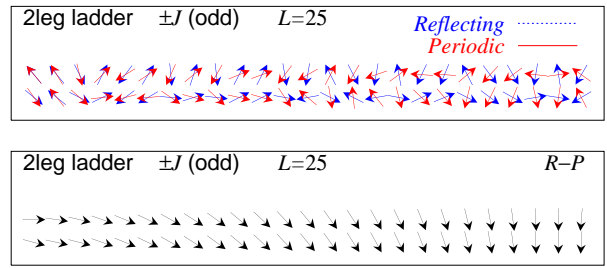}
\end{center}
\caption{
A typical example of the ground-state spin configuration of odd samples
under the periodic
and reflecting boundary conditions, between which a spin-wave of a turn 
angle $\pi/2$ is generated. The lattice size is $L=25$.
In the lower panel, the relative
deviation angle between the spin directions under 
the P and R BCs is illustrated by arrows.
}
\end{figure}

Here, 
the domain-wall energy,  $\Delta E_c$, is expected to be a sum of the chiral 
domain-wall contribution characterized by the chiral
exponent $y_\kappa =1.899\cdots $ and the SW contribution characterized by 
the SW exponent $y'_s=1$\cite{Ney}. 
For large enough $L$, a slowly-decaying
component, {\it i.e.\/}, the SW component, should dominate the asymptotic
behavior of $\Delta E_c$. In Fig.8,
we show on a log-log plot the $L$-dependence
of the domain-wall energy, $\Delta E_c$, calculated  by our 
methods (A), (B1)-(B3) and (C). 
In the methods (B1)-(B3), the position of a single chiral domain wall
is determined by calculating and comparing the energies corresponding to
all possible $N_{\rm fr}$ positions of a chiral domain wall.
In the size range $L\leq 35$ where the direct calculation is available,
the data  yield a slope about 1.07, which is slightly larger than the 
expected asymptotic value $y_\kappa =1$.
This deviation for smaller $L$ might be due to the residual 
contribution of the chiral domain-wall.
Meanwhile, 
the data for larger lattices obtained by the approximate methods 
(B1)-(B3) and (C) yield
the asymptotic slope fully consistent with the expected value, $y=1$. 

\begin{figure}[ht]
\begin{center}
\includegraphics[scale=0.5]{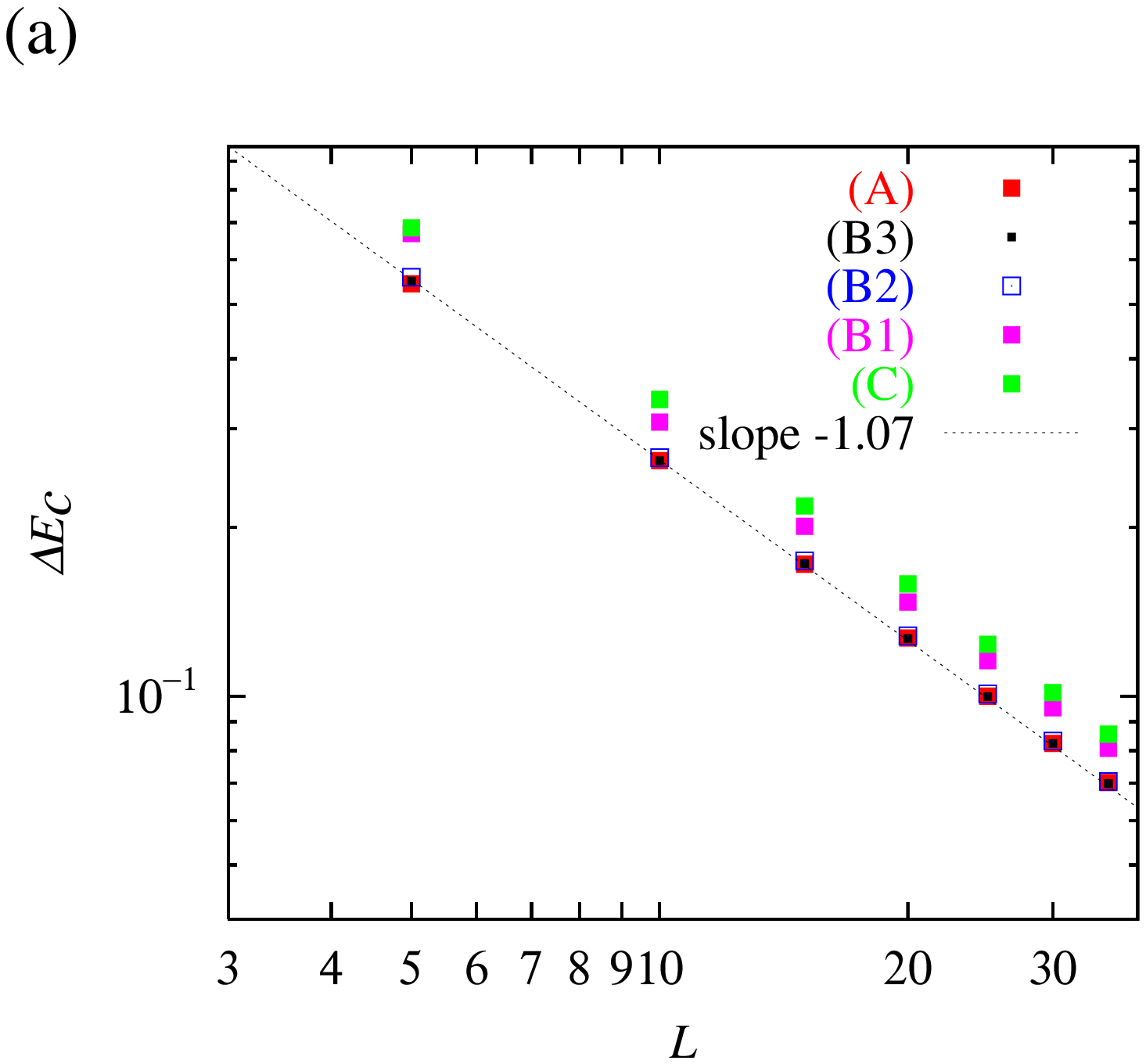}
\includegraphics[scale=0.5]{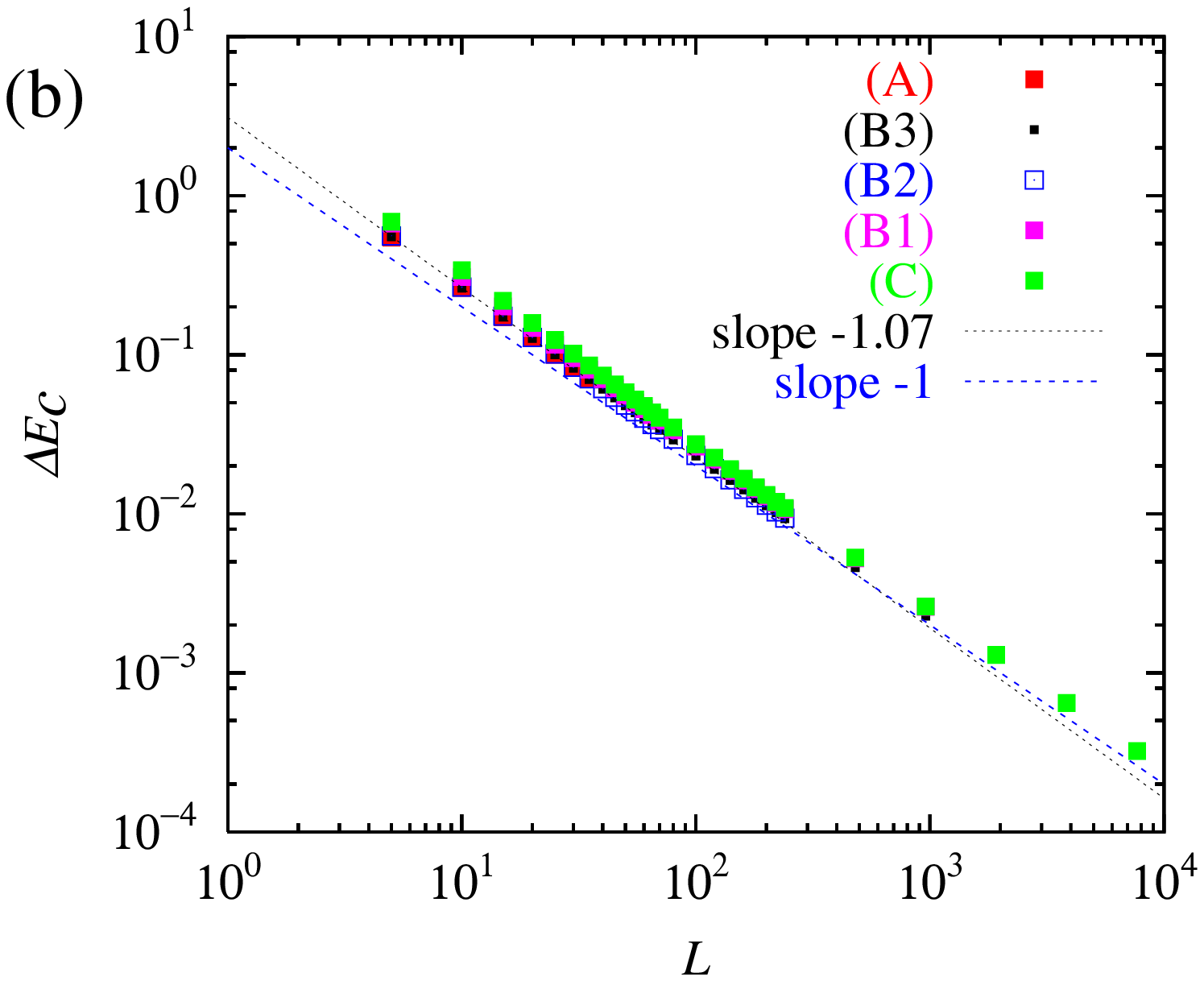}
\end{center}
\caption{
The domain-wall energy $\Delta E_c$ of odd samples,
calculated by the three
methods (A), (B1)-(B3) and (C) mentioned in the text, are
plotted versus $L$ on a log-log plot. 
}
\end{figure}

\bigskip
\bigskip
\noindent
3.3 All samples

As shown above, the behaviors of the domain-wall energy largely 
differ between in even and odd samples.
The behavior observed when one measures the domain-wall energy averaged 
over all samples
can be obtained immediately by simply combining the 
above results for even and odd samples with equal weights.

The behavior of $\Delta E_{{\rm P,AP}}$ 
averaged over all samples is exactly the same as that for 
even samples, since $\Delta E_{{\rm P,AP}}$ 
is identically zero 
for odd samples.
It exhibits an asymptotic behavior characterized by the stiffness exponent 
$y_\kappa =1.899\cdots $ for large enough $L$, 
while there is a significant finite-size correction and the asymptotic behavior sets in only at $L\gsim 40$.

By contrast, 
$\Delta E_c$ is a sum of the chiral contribution and the SW contribution,
the large-$L$ behavior being dominated by the latter. Hence, 
$\Delta E_c$ averaged over all samples is asymptotically characterized by the 
SW exponent $y'_s=1$, in contrast to $\Delta E_{{\rm P,AP}}$.

These asymptotic behaviors were just as
predicted by Ney-Nifle {\it et al\/} \cite{Ney}.

\bigskip
\bigskip
\noindent
\S 4. Numerical results of the correlation length
\medskip

In this section, we numerically investigate the temperature dependence of 
the spin and the chiral correlation lengths, $\xi_s$ and $\xi_\kappa$, with 
interest in the associated
correlation-length exponents, $\nu_s$ and $\nu_\kappa$. 
Since the correlation functions are bulk quantities, one usually believes that 
they are independent of the type of BCs, 
or whether the sample is either even or odd, 
in contrast to the case of the domain-wall energies 
analyzed in the previous section. Indeed, 
this was  implicit in the analysis of 
Ref.\cite{Ney}.

The two-point spin-spin and chirality-chirality 
correlation functions, $C_s(R)$ and $C_\kappa (R)$, are defined by
\begin{equation}
C_s(R)=[<\vec S_{(0,1)}\cdot \vec S_{(R,1)}>^2],
\end{equation}
\begin{equation}
C_\kappa(R)=[<\kappa_0\kappa _R>^2],
\end{equation}
where $<\cdots >$ denotes the thermal average and $[\cdots ]$ denotes the
average over the bond disorder.

We directly calculate these 
correlation functions of the cosine model under the P BC by means of
the standard Monte Carlo simulation. The spin and the chiral 
correlation lengths are extracted 
by fitting the calculated correlation functions by a 
simple exponential form, $A\exp[-(r/\xi)]$ ($A$ is a constant). 
To guarantee that the estimated correlation lengths are free from 
the finite-size effect, 
the data are limited to the temperature region $T/J \geq  0.1$ 
where both correlation lengths $\xi_s$ and $\xi_\kappa$ are much smaller than 
the system size $L=100$. 
At the temperature $T/J=0.1$, $\xi_s$ and $\xi_\kappa$
become around $6$ and $3$, respectively. 
In order to be sure that 
the correlation lengths are insensitive to whether the sample is even or odd, 
we estimate $\xi_s$ and $\xi_\kappa$ for each case of 
even and odd samples. As expected, 
$\xi_s$ of even and odd samples 
agree within the error bars, so does $\xi_\kappa$. 

In Fig.9(a), 
we show on a log-log plot the temperature dependence of 
$\xi_s$ and $\xi_\kappa$ averaged over all samples,
total number of samples being $100$. As can be seen from Fig.9(a), 
in the investigated temperature range $T/J\geq 0.1$,  the data
yield a slope close to unity 
for both $\xi_s$ and $\xi_\kappa$.
This value, unity, may be related to the SW stiffness exponent $y'_s=1$ 
via the relation $\nu=1/y$. It is
significantly smaller than the asymptotic value obtained in Ref.\cite{Ney}
$\nu_s=\nu_\kappa=1/y_\kappa=0.5263\cdots $. However,
since the temperature range covered in the present simulation is rather
high and the correlation lengths still stayed shorter than the crossover 
length  $\sim 40$ estimated in the previous section, it is quite probable
that we need to go to lower temperatures to see 
the true asymptotic critical behavior associated with the $T=0$ transition.
Unfortunately, we cannot
directly evaluate the correlation lengths in this low temperature region 
because of the finite-size effect and the
thermalization problem.

\begin{figure}[ht]
\begin{center}
\includegraphics[scale=0.5]{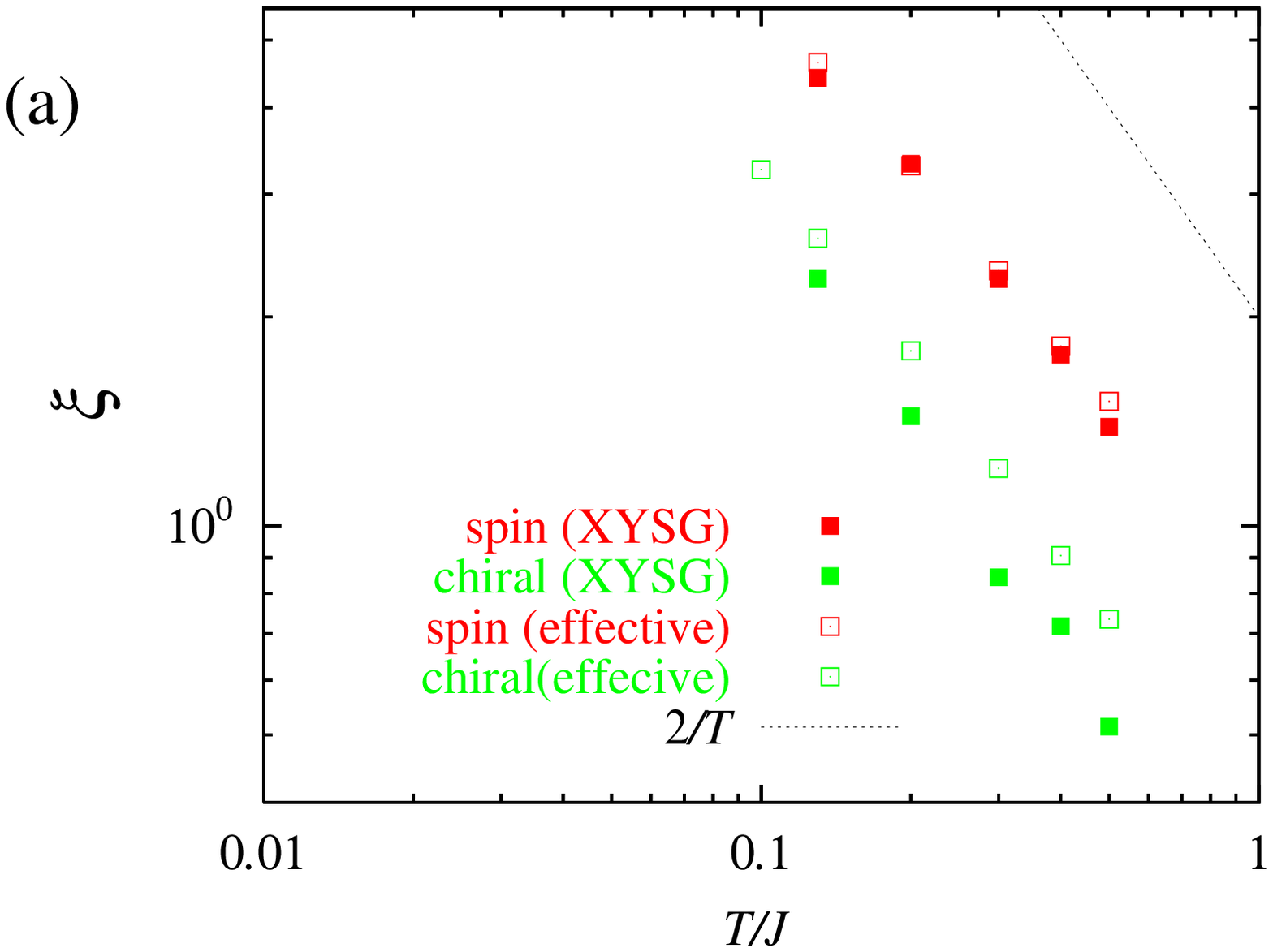}
\includegraphics[scale=0.5]{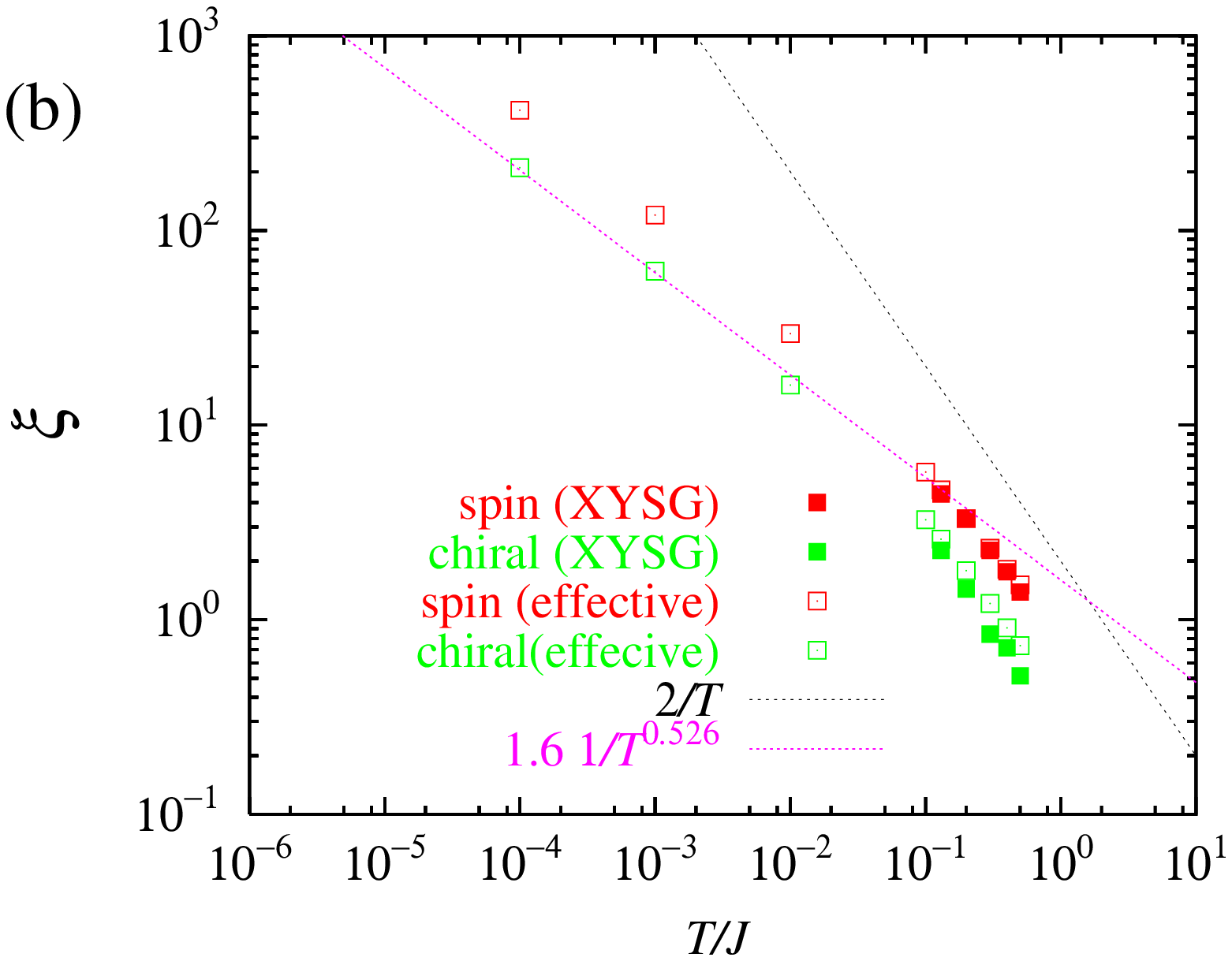}
\end{center}
\caption{
The temperature dependence of the spin and the chiral correlation lengths,
$\xi_s$ and $\xi_\kappa$, of the bulk sample, 
estimated by Monte Carlo simulation and numerical analysis of the
effective charge model.}
\end{figure}

Ney-Nifle {\it et al\/} gave an analytic expression of the spin and the
chiral correlation lengths in the thermodynamic limit on the basis of the two 
assumptions mentioned above \cite{Ney}. The chiral
 correlation function is obtained as
\begin{eqnarray}
C_\kappa(R) &=& [<2m_0\cdot 2m_R>^2] \\ 
&=&
\left[\prod_{j=1}^{S} \tanh^{2} \left(\frac{V_{j}}{k_BT}\right)\right] \label{eq-ckappa} \\
& = &
e^{-R/\xi_\kappa},
\end{eqnarray}
where $T$ is the temperature. In Eq. (\ref{eq-ckappa}), the product is
taken over all nearest-neighbor 
effective bonds $V_j$ which lie in the section between the sites $0$ and $R$
and are labeled as $j=1,2,\ldots,S$ where $S$ is the number of
frustrated plaquettes between the sites $0$ and $R$.
The chiral correlation length $\xi_\kappa$
behaves in the $T\rightarrow 0$ limit as
\begin{equation}
\xi_\kappa\sim \frac{1}{T^{\nu_\kappa}}\ \ \ \ \ 
\nu_\kappa=1/y_\kappa=0.5263\cdots .
\end{equation}
In the Villain model, the spin correlation function is obtained as
\begin{eqnarray}
C_s(R) &=&
[<\cos (\theta_{(0,1)}-\theta_{(R,1)})>^2] \\
&=&
e^{-R/\xi'_s} C_{\rm charge}(R), 
\end{eqnarray}
where the SW correlation length $\xi'_s$ is given by
\begin{equation}
\xi'_s = \frac{2}{T}.
\end{equation}
Here the spin correlation function is factorized into the
two parts, one  due to the SW and the other due to the charges 
$C_{\rm charge}$.
Within the effective model studied by Ney-Nifle {\it et al\/} \cite{Ney}
, {\it i.e.} the Ising model with the nearest-neighbor random antiferromagnetic
interaction, the charge part becomes,
\begin{eqnarray}
C_{\rm charge}(R) & = & \left[
\prod_{k=1}^{S/2} \tanh^{2} \left(\frac{V_{2k}}{k_BT}\right)\right] \label{eq-g}  \\
& = & e^{-R/2\xi_\kappa}.
\end{eqnarray}
In the cases where $S$ is an odd integer, the charge (chiral)
correlation function vanishes.
These results do not depend on the type of BCs nor on 
whether the sample is either even or odd. 

In Fig.9, we show these analytic results of the
spin and chiral correlation lengths for the effective model
together with the corresponding
MC estimates for the original {\it XY} SG model. 
In our analytic calculations, we evaluated the averages over the disorder
in Eq.(\ref{eq-ckappa})
and Eq.(\ref{eq-g}) taking into account
the true discrete spectrum of the
distribution of spacings between frustrated plaquettes rather than using
the continuous expression Eq.(\ref{eq-pv}).
As can be seen from Fig 9(a), in the higher 
temperature range $T/J\geq 0.1$, 
the analytic results of Ref.\cite{Ney} agree with our MC results, 
exhibiting the near $1/T$-behavior. 
Such an agreement observed
at higher temperatures
might give some credence to the reliability of the approximate methods. 
At lower temperatures where the 
MC result is no longer available, the analytic results of Ref.\cite{Ney}
tend to level off, exhibiting a clear crossover. 
There, for both cases of $\xi_s$ and $\xi_\kappa$,
a power-law behavior 
with a much smaller asymptotic exponent $\nu_s=\nu_\kappa =0.5263\cdots $ are 
eventually realized.
The crossover from the $1/T$ behavior to the $1/T^{0.5263..}$ behavior
occurs below $T/J\simeq 0.1$, at the length scale of $L=30$ lattice spacings. 
This crossover might be related to the
domain-wall result in the previous section where  
a crossover takes place at around $L=40$. 

Hence, the asymptotic critical behavior of the spin and chiral 
correlation lengths 
sets in only at low temperatures $T/J\lsim 0.1$ and 
at longer length scale $\xi \gsim 30$. At 
higher temperatures $T/J\gsim 0.1$ and at shorter length scale $\xi \lsim 30$, 
a different power-law behavior 
with an apparent exponent $\nu \simeq 1$ fits the data better.

It might be worth  emphasizing here again that,
although the spin and the chiral correlation lengths exhibit 
the same critical behavior, apparently with only one 
diverging length scale, there in fact exist two distinct length scales 
at the $T=0$ transition of this model. 
This has been already evident in Eqs. (17) and (20), where
the spin correlation function is written as a product of the $Z_2$ chiral part,
characterized by the correlation length with the chiral 
exponent $\nu_\kappa=0.5263\cdots $ and the $SO(2)$ SW part, 
characterized by the correlation length with the
SW exponent $\nu'_s=1$. Hence, while there actually exist the two diverging 
length scales at the $T=0$ transition of the model,
the one diverging more slowly, {\it i.e.\/}, the one with smaller 
$\nu$, dominate the asymptotic behavior of spin correlations, 
masking the existence of the other correlation length which diverges 
more rapidly. 

We note that such a ``masking'' phenomenon arises only when 
the inequality $\nu_\kappa < \nu'_s$ holds between the $Z_2$ and $SO(2)$ 
correlation-length exponents.
If this inequality would be opposite, the 
masking phenomenon would not show up in  the spin correlations. Then, the existence
of two correlation lengths would manifest itself more directly in the
associated correlation functions, 
the $SO(2)$ spin correlation length in the spin correlations  
and the $Z_2$ chiral correlation
length in the chiral correlations.  This actually occurs in the aforementioned 
regularly frustrated 1D $XY$ model, where one has 
$\nu_\kappa=\infty >\nu_s=1$
as shown rigorously for the case of 1D triangular lattice in
Ref. \cite{Horiguchi}. 

Naturally, one expects essentially the same behavior in the present
two-leg ladder {\it XY} model. Let us consider the regularly
frustrated two-leg ladder {\it XY} model such that all spacings
between frustrated plaquettes are equal to $l$. The analytic
expressions of the charge (chirality) and the spin correlation
functions, Eqs.(13) and (19), expected to be valid at low enough
temperatures $T/J \ll 1$, can also be used in the regular case where the
average over the bond disorder should be dropped. As examples, we show
in Fig.10 the temperature dependence of the spin and the chiral
correlation lengths $\xi_s$ and $\xi_\kappa$ for the cases of $l=2$ and
4. As can be seen from the figure, the chiral correlation length
$\xi_\kappa$ outgrows the spin correlation length $\xi_s$ at some
finite temperature $T_\times$. At higher temperatures $T>T_\times$,
$\xi_s$ and $\xi_\kappa$ exhibit a more or less similar behavior with
$\xi_s > \xi_\kappa$. At lower temperatures  $T<T_\times$, while 
$\xi_\kappa$ continues to exhibit a behavior
similar to the behavior observed at higher temperatures $T>T_\times$,
the spin correlation length $\xi_s$ dramatically changes its
behavior. The growth of $\xi_s$ with the decrease of the temperature is
dramatically slowed down below $T_\times$ yielding $\xi_s \ll \xi_\kappa$;
a manifestation of the spin-chirality decoupling.

\begin{figure}[ht]
\begin{center}
\includegraphics[scale=0.5]{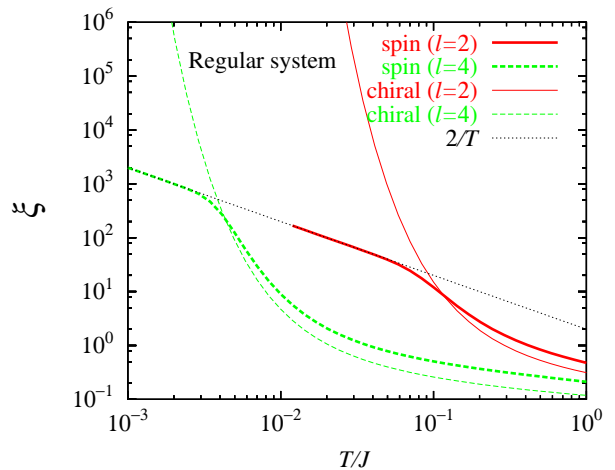}
\end{center}
\caption{The temperature dependence of the spin and the chiral correlation lengths,
$\xi_s$ and $\xi_\kappa$, of the two-leg ladder regularly frustrated $XY$ model
evaluated by the effective charge model. The spacings between frustrated 
plaquettes are equal to $l$. Here the results of $l=2$ and $l=4$ are shown as examples.}
\end{figure}

\bigskip
\bigskip\noindent
\S 5. Summary and discussion
\medskip

We numerically investigated the domain-wall energies 
and the spin and the chiral correlation lengths of the 1D $\pm J$ $XY$ SG ladder. 
Analytic results obtained by Ney-Nifle {\it et al\/} 
were confirmed for asymptotically large $L$, while the finite-size effect could
be significant. Concerning the domain-wall energies, 
the asymptotic behavior sets in only for lattices with $L\gsim 40$.
Concerning the correlation lengths,   
the asymptotic behavior sets in only at low temperatures $T/J\lsim 0.1$.

The domain-wall energies show different behaviors depending 
on the type of BCs and 
whether the number of frustrated plaquettes is even or odd. The domain-wall 
energy associated with the P/AP
BCs, $\Delta E_{{\rm P,AP}}$, is identically zero for odd samples,  
and exhibits a rather complex behavior for 
even samples. For even samples, although 
the lowest-energy excitation is always a chiral-domain wall pair
for asymptotically large $L$, 
the one for smaller $L$ is sometimes a 
$\pi$-SW without accompanying the chiral domain wall,
which eventually gives way to the chiral domain-wall pair excitation 
for larger $L$.
Somewhat unexpectedly, the approach to the asymptotic large-$L$ limit is 
rather slow and could be even non-monotonic (see Fig.4(b)), 
and the asymptotic large-$L$ behavior sets in
only at $L\gsim 40$.  The chiral domain wall and the SW bear the
associated  stiffness exponents, $y_\kappa=1.899\cdots $ 
and $y'_s=1$, respectively.
We note that, if the  inequality between $y_\kappa$ and $y'_s$ 
would be opposite, {\it i.e.\/}, if $y_\kappa < y'_s$, then 
$\Delta E_{{\rm P,AP}}$ would
be characterized by $y_s$, not by $y_\kappa$.

The domain-wall energy associated with the R/min(P,AP)
BC, $\Delta E_c$, is governed by a chiral domain-wall excitation, with and 
without the SW for odd and even samples.
Absence or presence of the SW excitation gives rise to the different
asymptotic behaviors of $\Delta E_c$ for even and
odd samples, respectively. 
The SW excitation, whenever it is induced, governs the large-$L$
asymptotic behavior of the
domain wall energy since $y'_s=1<y_\kappa =1.899\cdots $ in the present model. 
Thus, $\Delta E_c$  is characterized by an asymptotic 
stiffness exponent $y=y_\kappa$ and $y=1$ for even and odd samples, 
respectively. We emphasize again that, if the  inequality 
between $y_\kappa$ and $y'_s$ would be opposite,  $\Delta E_c$ would always
be characterized by the chiral exponent $y_\kappa$ for both even and
odd samples.

We also numerically investigated the behavior of the spin and the chiral 
correlation lengths. Both exhibit the divergence characterized by
the chiral exponent $\nu_\kappa  =1/y_\kappa=0.5263\cdots$, 
whereas this asymptotic
behavior sets in only at low temperatures $T/J\lsim 0.1$ and at longer length 
scale $\xi \gsim 30$.  At
higher temperatures $T/J\gsim 0.1$ and at shorter length scale $\xi \lsim 30$, 
a different power-law behavior
characterized by an apparent exponent $\nu \simeq 1$ is realized.
Although both the spin and the chiral correlation lengths exhibit the  
critical behavior with a common
exponent $y_\kappa$, the system in fact possesses
two distinct diverging length scales: one associated with the $SO(2)$ SW  
and the other associated with the $Z_2$ chirality.
In that sense, the $SO(2)$ part and the $Z_2$ part are decoupled in this model. Reflecting the inequality $\nu'_s=1/y'_s=1 > \nu_\kappa =1/y_\kappa=
0.5263\cdots $, however, the spin correlation 
is dominated by 
the chiral exponent $\nu_\kappa$, not by
the SW exponent $\nu'_s=1$.
{\it The inherent spin-chirality decoupling  of the present model is then 
masked\/}.

These behaviors of the domain-wall energies and of the correlation lengths of 
the 1D $\pm J$ $XY$ SG ladder are seemingly at odds with the behaviors of the
corresponding 
2D and 3D models suggested in Refs \cite{KawaTane91,Kawa95}. 
In the 2D and 3D $XY$
SG, 
Refs. (\cite{KawaTane91}) conjectured that  $\Delta E_{{\rm P,AP}}$
was governed by the $SO(2)$ spin exponent,  
while $\Delta E_c$ was governed by the $Z_2$ chiral exponent.
This should be contrasted with the behavior of the present 1D model 
where  $\Delta E_{{\rm P,AP}}$
was governed by the $Z_2$ chiral exponent  while $\Delta E_c$ was governed 
by the $SO(2)$ spin (SW) exponent.
As argued in the previous section, however, this apparent difference 
is simply the consequence of 
the difference in the 
relative magnitude of the $Z_2$
and $SO(2)$  exponents between the two models.
Indeed, in the 2D case, Ref.\cite{KawaTane91} estimated 
$y_\kappa\simeq 0.5$ and $y_s\simeq 1.0$ from the $L$-dependence of  
$\Delta E_c$ and  $\Delta E_{{\rm P,AP}}$, respectively,
where one has $y_\kappa <  y_s$ in contrast to the present 1D case.  
If  the inequality between 
the two stiffness exponents  $y_\kappa $ and $y'_s$ 
were opposite  in the present 1D model, 
{\it i.e.\/}, $y_\kappa < y'_s$,  
essentially the same behavior as that in the 2D and 3D models, {\it
i.e.\/},
$\Delta E_c\approx L^{-y'_s}$ and  
$\Delta E_{{\rm P,AP}}\approx L^{-y_\kappa}$, 
would arise. Hence, the apparent large deviation
from Ref.\cite{KawaTane91} is merely due to the difference in the relative 
magnitudes of the two stiffness (and correlation-length) exponents.  

In this connection, a significant difference between the present 1D ladder
model and the higher-dimensional models 
might be that, while the effective interaction between
the charge variables $m_i$ is short-ranged in 1D, that in higher dimensions is 
{\it long-ranged\/}. This assigns a  non-trivial character to the
charged excitation in the higher dimensional models. 
In 2D, this charged excitation  could be viewed as a vortex. 
(In 3D, it is a vortex-line.) 
The important characteristic of such a vortex excitation is that 
it breaks the charge neutrality condition and interacts with
each other via the long-ranged Coulombic interaction. 
In the present 1D ladder model, even  though there certainly
exists an excitation which
apparently breaks the
charge-neutrality condition, the associated interaction between them
is always short-ranged, and cannot be regarded as a genuinely-charged 
vortex-like excitation. 

Thus, in higher dimensions, 
excitations associated with the charge excitation  
are at least of two distinct types: The one is 
the vortex excitation which breaks the charge-neutrality, and
the second one is the ordinary chiral excitation which 
preserve the charge-neutrality. In 2D, the former one is a point defect 
(vortex), while
the latter is a line defect (domain-wall). 
Note that 
the former vortex-like excitation exists in both frustrated and unfrustrated 
$XY$ spin models, often 
playing a vital role in the order-disorder process, 
while the latter chiral excitation is peculiar to the {\it frustrated\/} 
$XY$ spin models.
We expect that, in higher dimensions,  this genuinely-charged  
vortex excitation might give rise to the another diverging length
scale which predominantly disorder the spin (not the chiral) order. 
A possible conjecture would be that the
stiffness exponent associated with the vortex excitation $y_v$ takes a 
value smaller than 
the SW stiffness exponent or the chiral stiffness exponent, and is
more effective in disordering the spin.  
If this is really the case, the spin correlation length and 
the domain-wall energy $\Delta E_{{\rm P,AP}}$
would be governed by the non-chiral exponent $y_v$ associated with the 
genuinely-charged excitation. 
In the 2D $XY$ SG with the $\pm J$ interaction, 
the numerical estimate of Refs.\cite{KawaTane91} 
suggested $y_v\simeq 1$,
which appears to be larger than $y_\kappa \simeq 0.5$.
 
We note that, even in 1D, the long-ranged interaction between  the charge 
variables could arise in some special cases, {\it e.g.\/}, 
in the tube lattice  investigated by Hill {\it et al\/} \cite{Thill}. 
In the tube lattice, however, 
the application of neither the P, AP nor the R BCs is capable
of generating the genuinely-charged vortex excitation, {\it i.e.\/},  
the excitation in the $q^{+}$ variable in the notation of Ref.\cite{Thill}. 
In fact, the application of either
the P, AP and R BCs generates only the charge-neutral chiral excitation, 
the excitation in the $q^{-}$ variable in the notation of Ref.\cite{Thill},
which interacts only via the short-ranged interaction. As such, 
the role of the genuinely-charged vortex excitation still remains to be seen in
the 1D tube model.
In order to elucidate the 
nature of the spin and the chirality orderings of the frustrated
{\it XY} spin systems, it would be important to further clarify the role 
of the vortex-like genuinely-charged 
excitation, not only in the 1D tube model but also
in the higher-dimensional models.

In the present paper, we concentrated on the {\it XY} SG ladder with the 
$\pm J$ interaction. The corresponding model with the Gaussian interaction
was also studied in the literature, {\it e.g.\/}, in Ref.\cite{Morris}. Our 
numerical study suggested that these two models, {\it i.e.\/}, the
1D {\it XY} SG ladder with either the $\pm J$ or the Gaussian interaction
exhibit quite different behaviors. The properties of the 1D 
{\it XY} SG ladder with the Gaussian interaction will be reported 
elsewhere \cite{Gaussian}.

\appendix

\section{Appendix: Effective charge Hamiltonian}

In this appendix, we briefly summarize the derivation of the effective charge Hamiltonian
corresponding to the original {\it XY\/} spin-glass model described by the 
spin Hamiltonian Eq. (\ref{eq-cos-hamiltonian}). We note that the
Villain's Hamiltonian, which contains the two-body interaction between
the charges only, is not exact even in the $T \to 0$ limit, contrary to a common belief. 
In the following, we give
explicit forms of the correction terms to the Villain's
approximation, which includes
certain $4$, $6$,...body effective interactions between the
charges. Detailes will be reported elsewhere wtihin a more
general context \cite{unpublished}.

As usual, the starting point is the identity
\begin{equation}
e^{\beta \cos \theta}=\sum_{p=-\infty}^{\infty} e^{ip\theta} I_{p}(\beta), 
\end{equation}
where $\beta=1/T$ is the inverse temperature and $I_{p}(\beta)$ 
is a modifeid Bessel function.
Now we use a useful formula for the assymptotic behaviour 
in the limit $\beta \to \infty$,
\begin{equation}
\log \left( I_{p}(\beta)/\frac{e^{\beta}}{\sqrt{2\pi \beta}} \right) 
=\sum_{m=1}^{\infty}
\frac{c_{m}}{m!} T^{2m-1} \left(  1 +  O(T) \right)  \mu^{m}.
\label{eq-log-bessel}
\end{equation}
where $\mu=4p^{2}$. Here terms which vanish in the $T \to 0$
are repsented as $O(T)$. 
The first few coefficients reads as
$c_{1}=-1/8$, $c_{2}=1/192$, $c_{3}=-3/2560$, $c_{4}=15/28672$, 
$c_{5}=-35/98304$, $c_{6}=945/2883584$,...
Note that the usual Villain's approximation amounts to assume that $c_{1}=-1/8$ 
and $c_{m}=0$ for $m >1 $. 

Following the standard steps of mapping of the original spin model 
to the charge model on the dual lattice \cite{Joseetal}, 
one finds that the effective charge Hamiltonian in the $T \to 0$ limit can be written as,
\begin{equation}
{\cal H}_{\rm charge}={\cal H}_{\rm 2-body}+{\cal H}_{\rm 4-body}
+{\cal H}_{\rm 6-body}...
\end{equation}
The terms on the r.h.s are due to the terms of $m=1,2,3,\ldots$ 
in Eqs. (\ref{eq-log-bessel}). 
The first term is nothing but the usual Villain's Hamiltonian 
which describes the two-body charge interactions,
\begin{equation}
{\cal H}^{\rm 2-body}=\sum_{i,j}U_{i,j} m_{i}m_{j}+
\frac{2 \pi^{2}}{L}(m^{2}_{\rm ex1}+m^{2}_{\rm ex1}), 
\end{equation}
where $U_{i,j}$ is given by Eq. (\ref{eq-uij}). The second term on the r.h.s
describes global spin-wave excitations induced by  the
two external change variables 
 $m_{\rm ex 1}$ and $m_{\rm ex 2}$ noticed in \cite{Ney}. 
In the case of periodic (P)
and antiperiodic (AP) boundary conditions (BC), all charge variables
are subjected to the global neutrality condition
$\sum_{i}m_{i}+m_{\rm ex 1}+m_{\rm ex 2}=0$ due to the presence of
a massless mode  \cite{Ney}. 
In the notaion of \S 2, the external
charges reads as $m_{\rm ex 1}=n-\frac{{\cal P}}{2}$ for P BC
and $m_{\rm ex 1}=n-\frac{{\cal P}-1}{2}$ for AP BC while
$m_{\rm ex 2} =\sum_{i}m_{i}-m_{\rm ex1}$. 
One finds that the same global neutrality condition applies 
also for the higher-body terms given below for the cases of P and AP BCs. 
In the case of reflecting (R) boundary condition, on the other hand, 
one finds $m_{\rm ex1}=m_{\rm ex2}=0$ in the dual mapping 
and also finds that the massless mode and thus the global neurtarily
condition is absent.

The explicit form of the higher-body terms becomes increasingly
complicated. The 4-body Hamitonian reads as,
\begin{equation}
{\cal H}^{\rm 4-body}=-\frac{4^{2}c_{2}}{2!}\sum_{i}
\left[ D_{i}^{4} + (D_{\rm ex 1})_{i}^{4} +  (D_{\rm ex 2})_{i}^4 \right],
\end{equation}
with
\begin{equation}
D_{i} \equiv \sum_{j} \left( \frac{U_{i+1,j}}{\pi}
- \frac{U_{ij}}{\pi} \right) m_{j},
\end{equation}
\begin{equation}
(D_{\rm ex 1})_{i} \equiv  
- \frac{2\pi}{L}m_{\rm ex 1}
+\sum_{j} \frac{U_{ij}}{\pi} m_{j} ,
\end{equation}
and
\begin{equation}
(D_{\rm ex 2})_{i} \equiv - \frac{2\pi}{L}m_{\rm ex 2}
+\sum_{j} \frac{U_{ij}}{\pi} m_{j}.
\end{equation}

The 6-body Hamiltonian reads as
\begin{eqnarray}
&& {\cal H}^{\rm 6-body}  =  -\frac{4^{3}c_{3}}{3!}\sum_{i}
\left[ D_{i}^{6} + (D_{\rm ex 1})_{i}^{6} +  (D_{\rm ex 2})_{i}^{6} \right] \nonumber \\
&& +  \frac{1}{2}\left( \frac{2 \cdot 4^{2}c_{2}}{1!}  \right)^{2}
\left \{
\sum_{i,j}  A_{ij} D_{i}^{3} D_{j}^{3}  \right.  \nonumber \\
&&  \left. - \sum_{i,j} (A_{ij}+B_{ij}) D_{i}^{3} \left( (D_{\rm ex 1})^{3}_{i} + (D_{\rm ex 2})^{3}_{i} \right)
\right. 
\nonumber \\
&& +  \sum_{i,j}  C_{ij}\left((D_{\rm ex 1})^{3}_{i}- (D_{\rm ex 2})^{3}_{i} \right)
\left((D_{\rm ex 1})^{3}_{j}- (D_{\rm ex 2})^{3}_{j} \right) \nonumber \\
&& + \left. \frac{1}{L}\left[ \left(\sum_{i} (D_{\rm ex 1})^{3}_{i}\right)^{2} 
+ \left(\sum_{i} (D_{\rm ex 2})^{3}_{i}\right)^{2} \right] \right\}.
\end{eqnarray}
where
\begin{equation}
A_{ij}=\frac{1}{L}\sum _k \frac{\cos(k(i-j))(1-\cos(k))}{2-\cos k}.
\end{equation}
\begin{equation}
B_{ij}=\frac{1}{L}\sum _k \frac{\sin(k(i-j))\sin(k)}{2-\cos k}.
\end{equation}
\begin{equation}
C_{ij}=\frac{1}{L}\sum _k \frac{\cos(k(i-j))}{4-2\cos k}.
\end{equation}

For a simple demonstration, let us consider a special 2-leg ladder sample in which all plaquettes are unfrustrated. 
In this particular sample, one knows that a spin-wave is
induced under the AP boundary condition so that
\begin{equation}
\Delta E_{PAP}=-2L \left[\cos \left(\frac{\pi}{L}\right)-1\right]=\frac{\pi^{2}}{L}-\frac{1}{2!}
\frac{\pi^{4}}{L^{3}}+\frac{1}{6!}\frac{\pi^{6}}{L^{5}} + \ldots
\end{equation}
exactly. One can check that the first three terms in the last equation can be
obtained by the energies associated with the 2-body (Villain's 
approximation), the 4-body and the 6-body interactions, respectivly.


\begin{thebibliography}{99}

\bibitem{Banavar} J.R. Banavar and M. Cieplak, Phys. Rev. Letters {\bf 48}, 832 (1982).

\bibitem{McMillan} W.L. McMillan, Phys. Rev. B {\bf 29} 4026 (1984).

\bibitem{BrayMoore1} A.J. Bray and M.A. Moore, J. Phys. C {\bf 17}, L463 (1984).

\bibitem{Carter} A.C. Carter, A.J. Bray and M.A. Moore, Phys. Rev. Letters {\bf 88}, 077201 (2002).

\bibitem{Kawashima} N. Kawashima and T. Aoki, J. Phys. Soc. Jpn. {\bf 69}, Suppl.A, 169 (2000); N. Kawashima, J. Phys. Soc. Jpn. {\bf 69}, 987 (2000)

\bibitem{KawaAjiro} See, for example, H. Kawamura, in {\it Quantum Properties of Low-Dimensional Antiferromagnetss\/}, ed. Y. Ajiro and J.-P. Boucher (Kyushu University Press) p. 124 (cond-mat/0202109), and references cited therein.

\bibitem{Kosterlitz}
J.M. Kosterlitz and N. Akino, Phys. Rev. Lett. {\bf 82}  4094
(1999).

\bibitem{Granato2}
E. Granato, J. Magn. Magn. Mater. {\bf 226}, 364 (2001); 
Phys. Rev. B {\bf 69}, 144203 (2004); {\it ibid.\/}, 012503 (2004).

\bibitem{LeeYoung} L.W. Lee and A.P. Young, Phys. Rev. Letters {\bf 90},
227203 (2003).

\bibitem{KawaTane91}
H. Kawamura and M. Tanemura, J. Phys. Soc. Jpn. {\bf 60} (1991) 608.

\bibitem{Kawa95}
H. Kawamura, Phys. Rev. {\bf B} 51 (1995)12398.

\bibitem{KawaLi}
H. Kawamura and M.S. Li, Phys. Rev. Lett. {\bf 87},  187204 (2001).

\bibitem{Horiguchi} T. Horiguchi and T. Morita, J. Phys. Soc. Jpn. {\bf 59}, 888 (1990).

\bibitem{KawaTane87}
H. Kawamura and M. Tanemura, Phys. Rev. {\bf B} 36 (1987) 7177.

\bibitem{Batrouni}
G.G. Batrouni and E. Dagotto, Phys. Rev. {\bf B} 37, R9875 (1988).

\bibitem{Ray}
P. Ray and M. A. Moore, Phys. Rev. {\bf B} 45, 5361 (1992).

\bibitem{Bokil} 
H.S. Bokil and A.P. Young, 
J. Phys. A{\bf 29}, L89 (1996).

\bibitem{Wengel} 
C. Wengel and A.P. Young, 
Phys. Rev. B{\bf 56}, 5918 (1997).

\bibitem{Granato1}
E. Granato, Phys. Rev. B{\bf 58}, 11161 (1998); 
B{\bf 61}, 391 (2000).

\bibitem{Grempel}
J. Maucourt and D. R. Grempel, Phys. Rev. Lett. {\bf 80} (1998) 770.

\bibitem{Ney}
M. Ney-Nifle, H.J. Hilhorst and M.A. Moore,  Phys. Rev. B{\bf 48} 
(1993) 10254.

\bibitem{Thill}
M.J. Thill, M. Ney-Nifle and H.J. Hilhorst, J. Phys. A {\bf 28} 
(1995) 4285.

\bibitem{Morris} B.M. Morris, S.G. Colborne, M.A. Moore, A.J. Bray
and J. Canisius, 
J. Phys. C{\bf 19}, 1157 (1986).

\bibitem{Gaussian} T. Uda and H. Kawamura, unpublished.

\bibitem{Joseetal} J. ~V. Jos{\' e}, L.~P. Kadanoff, S. Kirkpatrick and
	D. ~R. Nelson,  Phys. Rev. B {\bf 16} 1217 (1977).

\bibitem{unpublished} H. Yoshino and H. Kawamura, unpublished.

\end{thebibliography}
\end{document}